\documentclass[3p]{elsarticle}
\usepackage{graphicx,amsmath}
\usepackage{multirow,url}
\usepackage{color,soul}
\usepackage{subfigure}
\usepackage[table]{xcolor}
\usepackage{amssymb}

\title{Spoofing Detection Goes Noisy: An Analysis of Synthetic Speech
Detection in the Presence of Additive Noise}

\author[uef,btu]{Cemal~Hanil\c{c}i\corref{cor1}}
\ead{cemal.hanilci@btu.edu.tr}
\author[uef]{Tomi~Kinnunen}
\ead{tkinnu@uef.fi}
\author[uef]{Md~Sahidullah}
\ead{sahid@uef.fi}
\author[uef]{Aleksandr~Sizov}
\ead{sizov@uef.fi}

\cortext[cor1]{Corresponding author}
\address[uef]{School Of Computing, University of Eastern Finland, Joensuu, Finland.}
\address[btu]{Department of Electrical-Electronic Engineering, Bursa Technical University, Bursa, Turkey.}

\begin{document}
\begin{abstract}
Automatic speaker verification (ASV) technology is recently finding its way to end-user applications for secure access to personal data, smart services or physical facilities. Similar to other biometric technologies, speaker verification is vulnerable to spoofing attacks where an attacker masquerades as a particular target speaker via impersonation, replay, text-to-speech (TTS) or voice conversion (VC) techniques to gain illegitimate access to the system. We focus on TTS and VC that represent the most flexible, high-end spoofing attacks. Most of the prior studies on synthesized or converted speech detection report their findings using high-quality clean recordings. Meanwhile, the performance of spoofing detectors in the presence of additive noise, an important consideration in practical ASV implementations, remains largely unknown. To this end, our study provides a comparative analysis of existing state-of-the-art, off-the-shelf synthetic speech detectors under additive noise contamination with a special focus on front-end processing that has been found critical. Our comparison includes eight acoustic feature sets, five related to spectral magnitude and three to spectral phase information. All the methods contain a number of internal control parameters. Except for feature post-processing steps (deltas and cepstral mean normalization) that we optimized for each method, we fix the internal control parameters to their default values based on literature, and compare all the variants using the exact same dimensionality and back-end system. In addition to the eight feature sets, we consider two alternative classifier back-ends: Gaussian mixture model (GMM) and i-vector, the latter with both cosine scoring and probabilistic linear discriminant analysis (PLDA) scoring. Our extensive analysis on the recent ASVspoof 2015 challenge provides new insights to the robustness of the spoofing detectors. Firstly, unlike in most other speech processing tasks, all the compared spoofing detectors break down even at relatively high signal-to-noise ratios (SNRs) and fail to generalize to noisy conditions even if performing excellently on clean data. This indicates both difficulty of the task, as well as potential to over-fit the methods easily. Secondly, speech enhancement pre-processing is not found helpful. Thirdly, GMM back-end generally outperforms the more involved i-vector back-end. Fourthly, concerning the compared features, the Mel-frequency cepstral coefficient (MFCC) and subband spectral centroid magnitude coefficient (SCMC) features perform the best on average though the winner method depends on SNR and noise type. Finally, a study with two score fusion strategies shows that combining different feature based systems improves recognition accuracy for known and unknown attacks in both clean and noisy conditions. In particular, simple score averaging fusion, as opposed to weighted fusion with logistic loss weight optimization, was found to work better, on average. For clean speech, it provides $88\%$ and $28\%$ relative improvements over the best standalone features for known and unknown spoofing techniques, respectively. If we consider the best score fusion of just two features, then RPS serves as a complementary agent to one of the magnitude features. To sum up, our study reveals a significant gap between the performance of state-of-the-art spoofing detectors between clean and noisy conditions.

\end{abstract}
\maketitle

\section{Introduction}
\emph{Automatic speaker verification} (ASV) \cite{ReynoldsGMMML} is the task of authenticating users based on their voices. Traditionally, ASV has mostly been applied in specialized surveillance and forensics applications but recent methodological advances have greatly increased interest in mass-market adoption to secure personal data. For instance, in 2013 a smartphone voice unlock feature was introduced to a Baidu-Lenovo phone\footnote{\url{http://www.signalprocessingsociety.org/technical-committees/list/sl-tc/spl-nl/2013-02/SpeakerVerificationMakesItsDebutinSmartphone}}, and similar activities are being pursued by Google to their Android devices\footnote{\url{http://thehackernews.com/2015/04/android-trusted-voice.html}}. Some of the favorable points of ASV over other popular biometric identifiers are wide applicability (no other sensors except microphone required), natural integration with face authentication in smartphones, as well as \emph{revocability}: if a voice token is compromised or stolen, another user pass-phrase can be selected.

A speech-based authentication system to control access to personal data or physical site will be useful only if it helps to improve the overall system security. A now well-recognized security concern with any biometric modality --- including fingerprints, face, and speech --- is that they are vulnerable to circumvention by \emph{spoofing attacks} \cite{JainBiometric}, whereby an attacker attempts to gain unauthorized access to the system by masquerading herself as another user. Attacks can naturally be executed at any parts of the system \cite{ratha2001-enhancing}, including software, biometric templates or features. However, \emph{direct} attacks, involving an injection of forged biometric data to the sensor or the transmission point, are arguably most accessible to even less technology-aware attackers. Consequently, direct spoofing attacks are under active research across all the major biometric modalities. Specific to ASV, four currently known types of direct attacks have been identified \cite{EvansBookChapter2014,ZhizhengSpeechComm2014}: (i) \emph{replay}  \cite{ReplayVillalba2,ElieReplay,ReplaySpeechComm}, representation of a pre-recorded target speaker utterance; (ii) \emph{impersonation} \cite{RosaMimicricy,Mimicricy2}, human-based mimicry of a target voice; (iii) \emph{text-to-speech synthesis} (TTS), artificially generated target voice from an arbitrary text input \cite{DeLeonICASSP2010}; and (iv) \emph{voice conversion} (VC), modification of source speech towards target speaker characteristics \cite{JinICASSP2008}.

In this study, we focus on VC and TTS as they are arguably more flexible and consistent for spoofing both text-independent and -dependent ASV systems \cite{ZhizhengSpeechComm2014}. The effectiveness of VC and TTS spoofing attacks were first demonstrated nearly two decades ago in \cite{HansenICASSP99} and \cite{MasukoEurospeech99}. Further recent studies \cite{MatroufICASSP2006, BonastreINTERSPEECH2007, DeLeonODYSSEY2010, EvansEUSIPCO2012, HagaiINTERSPEECH2013, ZhizhengAPSIPA2014, Wu2015-SAScorpus} affirm that even state-of-the-art ASV systems remain highly vulnerable to modern VC and TTS attacks. State-of-the-art VC and TTS can produce high-quality target speech using a relatively small amount of training data \cite{TodaVoiceConversionINTERSPEECH,YamagishSpeechSynthesisTASPL}. Even if implementing such attacks in practice would currently require a dedicated effort or special skill-set from the attacker, anytime in near future one should expect advanced voice transformation tools to be readily available for end-users in smartphones or other portable devices, thereby greatly increasing the threats imposed by advanced VC and TTS spoofing attacks.

Having recognized the vulnerability problem caused by spoofing attacks, a few first steps to develop various \emph{countermeasures} (CMs) have been taken \cite{ZhizhengSpeechComm2014}. The most common approach (for an exception, see \cite{JointIvector}) is to equip an off-the-shelf ASV system with a stand-alone spoofing attack detector module. In our case, a classifier that will assign a \emph{human} or \emph{synthetic} label (or a likelihood score) to a given utterance\footnote{For brevity, we use ``synthetic speech detection'' to refer to detection of both VC and TTS attacks. In the present context, such umbrella term is justified as TTS and VC systems often employ similar methods for voice coding}.

The novel contribution of this work, which is placed into a wider ASV context in Section \ref{eq:related-work}, is briefly stated as follows. We provide a detailed analysis on synthetic speech detection under acoustically degraded conditions, namely, additive noise, whose effects to spoofing detection are so far poorly understood. We do \emph{not} introduce new methods but analyse the state-of-the-art methods with respect to their potential robustness bottlenecks under as comparable parameter settings and evaluation data as possible. In specific, we adopt the now widely-adopted ASVspoof 2015 challenge data \cite{ASVspoofOverview} to our experiments, so as to assess the joint effect of varied attacks \emph{and} additive noise. By focusing on the key part of synthetic spoofing detectors, the feature extractor, our aim is to gain improved understanding on generalization capability of the feature extractors in this task. Our study, being the most comprehensive comparative analysis on the topic to date, is targeted especially for practitioners, such as ASV vendors, and researchers new to ASV spoofing research. The material throughout the manuscript is intended to be tutorial-like and as self-contained as possible.

\section{Related work, motivation and contributions}\label{eq:related-work}

\subsection{Methods for detecting synthetic speech}

Synthetic speech detection is enabled by imperfections of the VC or TTS systems. For instance, voice coders (vocoders) used for speech parameterization in VC and TTS systems use greatly simplified models of human voice production, such as all-pole synthesis filters driven by impulse train excitation \cite{SPTKToolkit}. Processing artifacts affect the spectral, temporal and prosody characteristics of synthetic speech. Similar to ASV, synthetic speech detectors consist of front-end (feature extraction) and back-end (classifier) components. Most of the work on synthetic speech detection focus on the former, including specific/tailored features combined with a simple Gaussian mixture model (GMM) or support vector machine (SVM) back-end. A substantially different approach, using standard MFCC features but focusing on i-vectors and advanced back-end modeling ideas, was carried out in \cite{JointIvector} with promising results on the voice-converted version of NIST 2006 SRE data (though not performing well on ASVspoof 2015 \cite{ClassifierComparison}).

In \cite{ZhizhengInterspeech2012}, standard Mel-frequency cepstral coefficients (MFCCs), cosine phase and modified group delay features were compared for the detection of Gaussian mixture model (GMM) and unit selection based synthetic speech, cosine phase features leading to the lowest error rates. In \cite{ZhizhengICASSP13}, MFCCs, modified group delay, phase, and amplitude modulation features were compared for detecting synthetic speech, the group delay features yielding the highest accuracy. One of the most popular feature sets used for synthetic speech detection are the so-called \emph{relative phase shift} (RPS) features \cite{DeLeonICASSP2011,RPSTIFS-2015,DeLeonTASL2012-RPS}. They are calculated based on the phase shift of the harmonic components of the signal with respect to fundamental frequency (F0), and have been reported to be effective in detecting synthetic speech \cite{RPSTIFS-2015, DeLeonTASL2012-RPS}. However, for instance \cite{DeLeonTASL2012-RPS} suggests that RPS-based synthetic speech detection might be sensitive to vocoder mismatch across training and test sets, leading to degraded performance. More recently in \cite{RPSTIFS-2015}, the RPS features were used to detect synthetic speech signals provided by Blizzard Challenge. The authors found out that RPS features outperformed MFCCs on detecting speech generated by statistical parametric speech synthesis whereas MFCCs yielded higher accuracy when synthetic signals were generated by unit selection, diphone or hybrid methods. Similar, inconsistent observations were found in our recent study \cite{FeatureComparison} where RPS features performed the best out of 17 compared feature extraction techniques when vocoders between training and test were matched, but yielded the highest error rates in the opposite case.

In \cite{WangSpoofingChallenge}, another robust phase-related feature similar to RPS, termed \emph{relative phase information} (RPI) \cite{RPIFeatures}, was used for synthetic speech detection using ASVspoof 2015 database. It was found to outperform both MFCC and MGD features. RPI processing aims at normalizing the phase changes resulting from frame positioning. In specific, with the aid of discrete Fourier transform (DFT), phase information is estimated relative to a \emph{fixed} base frequency ($\omega_b = 2\pi\times 1000$ Hz was used in \cite{WangSpoofingChallenge,RPIFeatures}) in contrast to the RPS representation that is based on sinusoidal modeling with phase shifts computed relative to estimated F0.

\subsection{Towards varied spoofing attacks: SAS corpus and ASVspoof 2015 challenge}

As the above review indicates, a large number of potentially useful methods to detect synthetic speech have been investigated. The \emph{user's dilemma}, however, is that their relative performances are either incomparable or under-representative of real-world deployment, for many reasons. Firstly, no single study compares the various methods on a common set of data or using a unified objective evaluation metric, making unbiased performance assessment challenging, if not impossible. Secondly, the studies usually contain only a handful of attacks, making conclusions attack-dependent. Thirdly, most studies involve a closed-world evaluation setting where the synthetic test samples originate from the same methods, channels and environments as used in training. This corresponds to a scenario where the ASV system administrator (defender) knows in advance what spoofing technique the attacker will employ. While such an oracle evaluation scenario may provide experimental bounds to the highest performance achievable using a specific attack detector, it is unlikely to be representative of an actual attack scenario where the attacker may employ novel (presently unknown) attacks. Fourthly, differently from the traditional NIST speaker verification scenarios involving channel- and condition-mismatched data, most of the datasets used for synthetic speech detection have consisted of high-quality (wideband) noise-free signals. As a result, it is largely unknown how well the state-of-the-art synthetic speech detectors generalize to non-ideal conditions involving not only varied spoofing materials but extrinsic distortions induced by the environment or channel, important factors in any real-world deployment of ASV technology.

To address the first three concerns --- incomparability of results, limited set of attacks and closed-world evaluation bias --- a new speaker verification spoofing and anti-spoofing (SAS) corpus was introduced recently in \cite{Wu2015-SAScorpus} and used in \emph{ASVspoof 2015: Automatic Speaker Verification Spoofing and Countermeasures Challenge} \cite{ASVspoofOverview}\footnote{\url{www.spoofingchallenge.org}}, that focused on stand-alone synthetic speech detection involving both known and unknown attacks. The findings of ASVspoof 2015 were disseminated at a special session of the latest edition of \emph{Interspeech} conference in Dresden, Germany\footnote{\url{http://www.signalprocessingsociety.org/technical-committees/list/sl-tc/spl-nl/2015-11/2015-11-ASVspoof/}}.

During the special session, several participating sites reported independently that spectral phase-based features (such as cosine phase \cite{ZhizhengInterspeech2012}, modified group delay \cite{ZhizhengInterspeech2012} and RPS \cite{RPSTIFS-2015}) outperformed spectral magnitude-based features in synthetic speech detection \cite{STCSubmission, VillalbaSpoofingChallenge, WangSpoofingChallenge, NTUSubmission}. GMM-based system \cite{ReynoldsGMMML} was used for modeling both natural and synthetic speech classes in most of the studies presented at the special session \cite{WangSpoofingChallenge, VillalbaSpoofingChallenge, AHOLABSubmission}. Though in \cite{VillalbaSpoofingChallenge}, more advanced support vector machines (SVM) and deep neural networks (DNN) are utilized as their back-ends, the performance of GMM systems was found to be similar or better. Similar observation was made in our preliminary study on ASVspoof 2015 data \cite{ClassifierComparison}.  An i-vector with Gaussian back-end and DNN based approach was also investigated in \cite{Zhang2016ICASSP} without improvement in performance compared to GMM. In most recent studies using ASVspoof 2015 data, fundamental frequency (F0) contour and strength of excitation (SoE) were also used in combination with MFCCs and \emph{cochlear filter cepstral coefficients and instantaneous frequency} (CFCCIF) features~\cite{Patel2016ICASSP}. In~\cite{Todisco2016Odyssey}, \emph{constant Q cepstral coefficient} (CQCC) was proposed for synthetic speech detection.

\subsection{Contribution of the present study: joint effect of varied attacks and noise}

In our two preliminary studies on ASVspoof 2015 data, we did extensive comparative evaluation of several front-end \cite{FeatureComparison} and back-end \cite{ClassifierComparison} synthetic speech detectors. In our experiments, the simplest ideas tended to outperform more elaborate ones. For instance, raw power spectrum features and maximum likelihood (ML) trained Gaussian mixture models (GMMs) did a decent job both in detecting both unknown and known attacks, while i-vector \cite{NajimIvector} based spoofing detection \cite{JointIvector, Elie-Interspeech2014} yielded much higher error rates.

The present study extends \cite{FeatureComparison} and \cite{ClassifierComparison} towards an extended and self-contained comparative evaluation of synthetic speech detectors. Unlike \cite{FeatureComparison} and \cite{ClassifierComparison}, where we used the original high-quality ASVspoof 2015 samples, in this study, we address the fourth concern missing from most of the prior studies: robustness of synthetic speech detection under acoustically degraded conditions. In general, an acoustic signal reaching a recognizer can be subjected to many extrinsic imperfections, induced by additive noise, transmission channel (including compression artifacts and low bandwidth), and reverberation, to name a few. A limited number of earlier studies have executed spoofing experiments on 8 kHz telephony data \cite{Wu2012, Elie-Interspeech2014}, though under somewhat artificial scenario in which an existing telephone-quality corpus has been post-processed through voice conversion attacks, as opposed to the more likely case of spoofing attacks taking place \emph{before} signal transmission. We argue that it is difficult, if not impossible, to isolate the relative impact of spoofing artifacts and extrinsic distortions without an access to the original, undistorted signal. Therefore, there is a clear need to examine spoofing attacks under \emph{controlled} extrinsic distortions to gain improved insight as to what might be the important considerations in developing practical countermeasures. A recent study \cite{Wester2015} addressed the impact of bandwidth to synthetic speech detection accuracy on the same ASVspoof 2015 corpus as used in the present study.

In contrast to the above prior studies, we focus solely on arguably one of the most common and relevant sources of distortions, additive noise. It has received almost no prior attention to the best of our knowledge\footnote{An independent study, made publicly available almost in parallel to ours \cite{NoisySpoofingDetection}, considers the same ASVspoof2015 database under additive noise contamination. Their noise contamination design is similar to ours though spoofing detection features are mostly different, and our manuscript provides a more thorough analysis.}. Specifically, using the ASVspoof 2015 corpus, we provide a detailed performance assessment of several spoofing detectors under additive noise contamination. Special attention is paid in making the compared methods as comparable as possible with respect to feature dimensionalities, frame rate and other control parameters.

\begin{figure}
\begin{centering}
\includegraphics[scale=0.35]{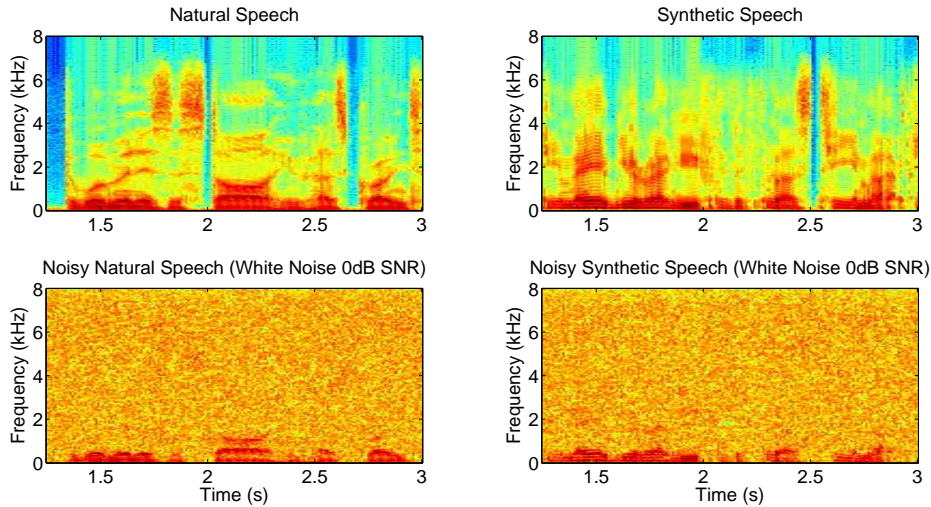}
\caption{Natural and synthetic speech signals of the same speaker and their noisy counterparts.}
\label{figure:nat_vs_synth_spectrogram}
\end{centering}
\end{figure}

We expect this to be a notoriously difficult task that could serve as a useful evaluation test-bench for developing new robust countermeasures more relevant for end-user applications. As state-of-the-art TTS and VC methods can produce high-quality speech, sometimes close to or indistinguishable from authentic human speech (\emph{unit selection} \cite{SundermannUnitSelection2006} is a good example), we expect additive noise to mask further the already small differences between human and synthetic speech. As a motivation, Fig.~\ref{figure:nat_vs_synth_spectrogram} displays spectrograms of natural and synthetic speech signals of the same speaker and their noisy counterparts. While differences of natural and synthetic speech are apparent for the clean data, additive noise makes it difficult to tell the difference.

It is not obvious, for instance, whether standard speech enhancement techniques as a pre-processing method will be helpful: as noise suppression is always traded-off with speech distortion \cite{SpringerHandbook2008}, processing artifacts due to speech enhancement could be confused with artifacts due to synthesis vocoders. Similarly, as indicated above, the popular RPS \cite{DeLeonTASL2012-RPS,RPSTIFS-2015} feature requires fundamental frequency tracking whose  performance is affected by additive noise \cite{RabinerPitchDetection}. For these reasons, it is not obvious what type of front-end or back-end modeling ideas will work comparatively better for synthetic speech detection under noisy conditions. To answer these questions, we have selected state-of-the-art or otherwise popular feature extraction methods based on both our preliminary results \cite{FeatureComparison} and those of the ASVspoof 2015 participants. Our eight feature sets, detailed below, include both magnitude- and phrase-related features. From the classifier side, we use GMMs trained via maximum likelihood (ML), reported as the best-performing one in \cite{ClassifierComparison}, as well as the i-vector approach \cite{Elie-Interspeech2014, JointIvector}.

\section{Spoofing Detection}
\label{sec:spoof_detect}
Given a speech signal $s$, synthetic speech detection task is to decide whether $s$ belongs to a natural speech class --- hypothesis $\cal{H}_\mathrm{0}$, or a synthetic speech class --- hypothesis $\cal{H}_\mathrm{1}$. The decision is based upon  the log-likelihood ratio score, $\Lambda$:

\begin{equation}
\Lambda(s) = \mathrm{\log{p(s | \cal{H}_\mathrm{0})}} - \mathrm{\log{p(s | \cal{H}_\mathrm{1})}}.
\label{eq:LLRscore}
\end{equation}

To estimate the probabilities $\mathrm{p(s | \cal{H}_\mathrm{0})}$ and $\mathrm{p(s | \cal{H}_\mathrm{1})}$ we need to train an acoustic model for each hypothesis. In our recent anti-spoofing study on ASVspoof 2015 \cite{ClassifierComparison}, we evaluated a number of different classification techniques. Gaussian mixture models (GMM), trained with maximum likelihood (ML) principle, was found the best choice.

GMM is a well-known probabilistic model that is extensively used for speaker recognition ever since it was introduced for the task \cite{ReynoldsGMMML}. We separately use natural and synthetic training data to train two GMMs. Each GMM consists of a  mixture weight $w_i$, a mean vector $\boldsymbol{\mu}_i$ and a covariance matrix $\mathrm{\mathbf{\Sigma}}_i$ for each mixture component $i$. We use expectation-maximization (EM) algorithm to estimate the model parameters $\lambda=\{w_i,\:\boldsymbol{\mu}_i,\:\mathrm{\mathbf{\Sigma}}_i\}_{i=1}^M$, where $M$ is the number of mixture components.

After the two acoustical models are trained, the log-likelihood for each hypothesis and a sequence of feature vectors $\mathrm{\mathbf{X}}=\{\mathrm{\mathbf{x}_1},\:\ldots \mathrm{\mathbf{x}_T} \}$, that represent the speech signal $s$, takes the following form

$$ \mathrm{\log{p(s | \cal{H}_\mathrm{k})}} = \dfrac{1}{T} \,\mathrm{\log{p(\{\mathrm{\mathbf{x}_1},\:\ldots \mathrm{\mathbf{x}_T} \} | \lambda_k)}} = \dfrac{1}{T} \, \sum_{t=1}^T{\log{(\mathrm{\mathbf{x}}_t|\lambda_k)}}.$$

Besides GMM, we also consider the i-vector paradigm~\cite{NajimIvector}, that became state-of-the-art technique for text-independent speaker verification. Recently, it was also used to perform speaker verification and anti-spoofing jointly in the i-vector space~\cite{JointIvector}. In essence, i-vector $\mathrm{\mathbf{w}}$ is a fixed-sized low-dimensional vector per utterance that contains both speaker- and channel-specific variability. To extract an i-vector, we factorize a GMM mean supervector $\boldsymbol{\mu}$ as $\boldsymbol{\mu}=\mathrm{\mathbf{m}}+\mathrm{\mathbf{T}}\mathrm{\mathbf{w}}$, where $\mathrm{\mathbf{T}}$ is a low-rank rectangular matrix, $\mathrm{\mathbf{m}}$ is a speaker-independent mean vector and $\mathrm{\mathbf{w}}$ has a standard normal prior distribution. Refer to \cite{NajimIvector} for more details.

We use two different i-vector based classifiers to compute the final score \eqref{eq:LLRscore}: \emph{cosine similarity measure} and \emph{probabilistic linear discriminant analysis} (PLDA) \cite{PLDA_Original}. Given two i-vectors, extracted from target ($\mathrm{\mathbf{w}}_{\mathrm{tgt}}$) and test ($\mathrm{\mathbf{w}}_{\mathrm{tst}}$) utterances, we compute cosine similarity between them using

\begin{equation}
\label{cosine_scoring}
\mathrm{cosine} (\mathrm{\mathbf{w}}_{\mathrm{tgt}}, \mathrm{\mathbf{w}}_{\mathrm{tst}})= \frac{\mathrm{\mathbf{w}}_{\mathrm{tgt}}^{\mathrm{T}}\mathrm{\mathbf{w}}_{\mathrm{tst}}}{\|\mathrm{\mathbf{w}}_{\mathrm{tgt}} \| \|\mathrm{\mathbf{w}}_{\mathrm{tst}} \|}.
\end{equation}

As the cosine similarity measure does not compute likelihoods, instead of Eq. \eqref{eq:LLRscore} we form the detection score as follows:

\begin{equation}
\label{score_ivector}
\mathrm{score} = \mathrm{cosine}(\hat{\mathrm{\mathbf{w}}}_{\mathrm{nat}},\mathrm{\mathbf{w}}_{\mathrm{tst}}) - \mathrm{cosine}(\hat{\mathrm{\mathbf{w}}}_{\mathrm{synth}},\mathrm{\mathbf{w}}_{\mathrm{tst}}),
\end{equation}
where $\hat{\mathrm{\mathbf{w}}}_{\mathrm{nat}}$ and $\hat{\mathrm{\mathbf{w}}}_{\mathrm{synth}}$ represent the average training i-vectors for natural and synthetic speech classes, respectively.

Besides cosine scoring, we also consider the so-called simplified PLDA \cite{Kenny2010}. The idea behind PLDA is to split total i-vector variability into speaker and channel components, which allows efficient inference during a test stage.
To train the model, we grouped together i-vectors from each synthesis method and from a natural speech which gave us 6 classes (``speakers"). For more details on the data, refer to Section~\ref{database_section}.
\section{Natural v.s. Synthetic/Converted Speech}
\begin{figure}
\begin{centering}
\includegraphics[scale=0.37]{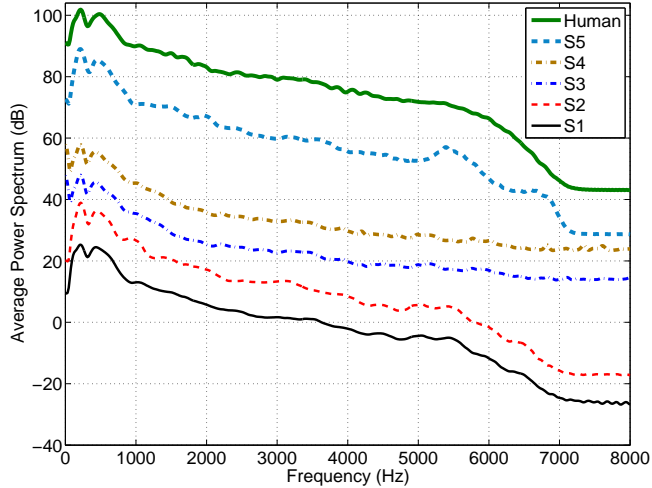}
\caption{Long-term average power spectra of synthetic and human speech signals (we used 2525 speech files per each method to compute an average). The spectra have been shifted by 10 dB with respect to each other.}
\label{LTAS_speech_figure}
\end{centering}
\end{figure}
Before proceeding to recognition experiments, we first wish to understand the acoustic signal properties of the natural and synthetic speech signals. To analyze the characteristics of natural and synthetic speech, long-term average spectra (LTAS) is utilized. LTAS somewhat represents the physical characteristics of the speaker related the vocal tract resonances \cite{LTAS_vocal_tract} and is mostly used in audio forensics \cite{LTAS_Forensics} and for measuring the audibility of speech to compute speech intelligibility index \cite{LTAS}. LTAS is computed by time averaging the short-term Fourier magnitude spectra of all frames:
\begin{equation}
\label{LTAS_eq}
\mathrm{LTAS}(k) = \frac{1}{T}\sum_{t=1}^{T}|S_t(k)|^2,
\end{equation}
where $S_t(k)$ denotes the windowed discrete Fourier transform of $t$th speech frame of the signal, $s$, at DFT bin $k$ and $T$ is the total number of speech frames after voice activity detection (VAD). We compute the average LTAS of human and synthetic speech signals using the training portion of the ASVspoof 2015 dataset for each synthesis/conversion technique (S1-S5) to visualize their differences in frequency domain. Fig.~\ref{LTAS_speech_figure} displays the LTAS computed using synthetic and natural speech signals (average LTAS is computed using $2525$ speech files per method). Synthetic speech power is attenuated below $4$ kHz compared to natural speech. For $f>4$ kHz, the opposite happens and the difference between human and synthetic speech signals are larger. Especially for S3 and S4, hidden Markov model (HMM)-based speech synthesis techniques, the relative difference between human and synthetic speech are higher than for the other synthesis/conversion techniques. Interestingly, when $f>7$ kHz, larger differences occur between other conversion techniques and natural speech.

It is well known that additive noise drastically reduces the speaker, language and speech recognition performances. Several methods to cope with the adverse effects of additive noise contamination have been proposed. Speech enhancement techniques aim to improve the quality of the signal corrupted by noise in the signal level. Cepstral mean subtraction (CMS) \cite{Atal_CMS}, cepstral mean and variance normalization (CMVN) and RASTA filtering \cite{Rasta} are the popular feature level methods to suppress linear channal bias in cepstral features, often yielding increased speaker recognition accuracy. Speaker, language and speech recognition under additive noise and mismatched channel conditions are well-studied and several techniques have been proposed to improve the performance. However, since spoofing detection has only recently been drawn attention, its performance under degradation and possible solutions for mismatched conditions are unknown. Thus, a thorough analysis on the effect of noise is necessary for the anti-spoofing research.

\begin{figure}
\begin{centering}
\includegraphics[scale=0.35]{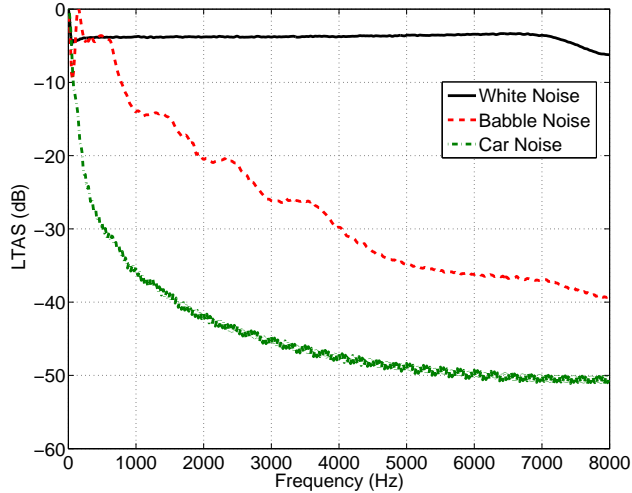}
\caption{Long-term average power spectra of different noise types used in the experiments.}
\label{LTAS_noise_figure}
\end{centering}
\end{figure}

In this study, we consider three noise types: (i) white noise, (ii) babble noise and (iii) car noise. The LTAS variations of each noise type are shown in Fig.~\ref{LTAS_noise_figure}.

\section{Feature Extraction Methods}\label{sec:feature-extraction-methods}
Speech features representing short-term spectral features, which are mostly used for speech and speaker recognition, are also employed in speech-based spoofing detection. A comparative evaluation of a large number of speech features for this task is available in \cite{FeatureComparison}. In this paper, we focus on the most promising (or otherwise popular) features for noise-robust spoofing detection, namely, mel-frequency cepstral coefficients (MFCCs), inverted mel-frequency cepstral coefficients (IMFCCs) \cite{chakroborty2007improved}, spectral centroid magnitude coefficients (SCMCs) \cite{SpectralMagnitude}, recently proposed constant Q cepstral coefficients (CQCC) \cite{Todisco2016Odyssey} and relative phase shift (RPS) \cite{DeLeonICASSP2011,DeLeonTASL2012-RPS,RPSTIFS-2015}, modified group delay (MGD) \cite{Murthy2003ICASSP} and cosine phase (CosPhase) \cite{ZhizhengInterspeech2012} features. MFCC and IMFCC are based on filter bank analysis, SCMC contains detailed information of subband while RPS, MGD and CosPhase carry phase-related information. In addition to magnitude and phase based features, we also evaluate recently proposed mean Hilbert envelope coefficient (MHEC) feature used successfully for robust speaker and language recognition \cite{Sadjadi2015_MHEC}.

The features and their parameters used in this study are summarized in Table~\ref{table:features_table}. All the features have been made as comparable as possible: their frame rates, DFT sizes, number of filters and dimensionality are the same (where applicable). Feature post-processing techniques (none or deltas followed by cepstral mean subtraction) were optimized for each feature set separately. In the following, we briefly describe each of the features.

\begin{table}
\centering
 \begin{small}
 \caption{Summary of the features and their parameters used in this study. Check marks represents corresponding post processing is applied whereas empty entries correspond to opposite.}
 \label{table:features_table}
\medskip
\footnotesize
 \begin{tabular}{lcc||ccc||c||ccc}
 \noalign{\hrule height 0.75pt}
 \multirow{2}{*}{Features} & \multirow{2}{*}{Frame length/shift} & \# DFT & \multicolumn{3}{c||}{Filters} & \multirow{2}{*}{Coefficients} & \multicolumn{3}{c}{Post Processing} \\
 & &  bins & $\#$ & Type & Scale & & $\Delta$ & $\Delta\Delta$ & CMS \\
 \noalign{\hrule height 0.75pt}
 MFCC & 20 ms/10 ms  & 512 & 32 & Triangular & Mel & $c_0-c_{31}$ & \checkmark & \checkmark & \checkmark \\
 IMFCC & 20 ms/10 ms  & 512  & 32 & Triangular & Mel & $c_0-c_{31}$ & \checkmark & \checkmark & \checkmark \\
 SCMC & 20 ms/10 ms  & 512 & 32 & Rectangular & Linear & $c_0-c_{31}$ & \checkmark & \checkmark & \checkmark \\
 MHEC & 20 ms/10 ms  & -  & 32 & Gammatone & ERB & $c_0-c_{31}$ & \checkmark & \checkmark & \checkmark\\
 RPS & 20 ms/10 ms  & 512 & 32 & Triangular & Mel & $c_0-c_{31}$ &  &  &  \\
 MGD & 20 ms/10 ms & 512 & - & - & - & $c_0-c_{31}$ & \checkmark & \checkmark & \checkmark \\
 CosPhase & 20 ms/10 ms & 512 & - & - & - & $c_0-c_{31}$ &  &  &  \\
 \noalign{\hrule height 0.75pt}
 \end{tabular}
 \end{small}
\end{table}

\subsection{Mel-frequency Cepstral Coefficients (MFCCs)}
In short-term speech processing, the speech signal is first divided into short overlapping frames (here $20$ ms frames with $10$ ms overlap is used). Then, the power spectrum of each Hamming windowed frame is computed using discrete Fourier transform (DFT) by
\begin{equation}
\label{power_spectrum}
|X[k]|^2 = \left|\sum_{n=0}^{N-1}x[n]e^{-j2\pi kn/N}\right|^2 \:\:\:0\leq k\leq K-1,
\end{equation}
where, $k$ is the DFT bin and $\mathbf{x}=\left[ x[0],\:\ldots,\:x[N-1]\right]$ is a windowed speech frame (assumed to be zero outside of the interval $[0,\:N-1]$). In standard filterbank based feature extraction schemes, the power spectrum is processed using a set of overlapping band-pass filters. Logarithmic filter bank outputs are then converted into cepstral coefficients by applying discrete Cosine transform (DCT). Generally, triangular filters spaced in mel-scale are used as filterbank and the resulting features are the mel-frequency cepstral coefficients (MFCCs).

\subsection{Inverted Mel-frequency Cepstral Coefficients (IMFCCs)}
\begin{figure}
\begin{centering}
\includegraphics[scale=0.3]{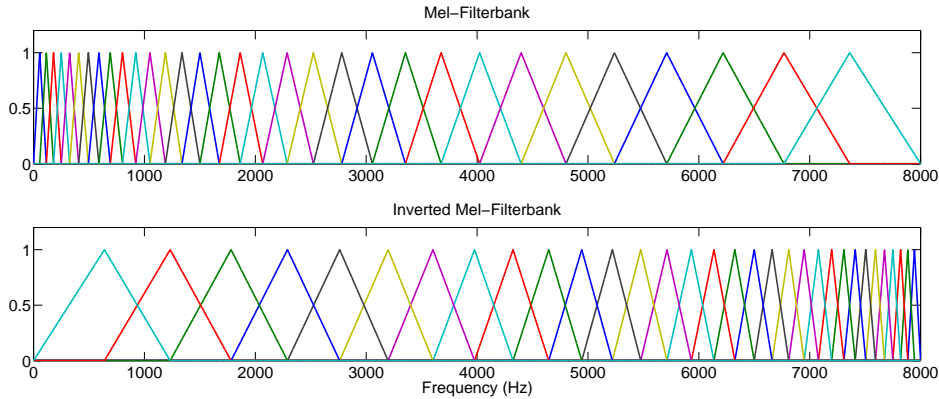}
\caption{Triangular filters spaced on Mel and inverted-Mel scale.}
\label{MelIMel_filterbank_figure}
\end{centering}
\end{figure}

In MFCCs, filters have denser spacing in low-frequency region. The IMFCC features are extracted using an \emph{inverted} Mel scale \cite{chakroborty2007improved}, implemented in practice by flipping the Mel-scaled filter bank in frequency axis giving more emphasis on the high-frequency region. Fig.~\ref{MelIMel_filterbank_figure} shows an example of triangular filters spaced on Mel and inverted Mel scales. Otherwise, all the processing steps remain the same as in MFCC extraction.

\subsection{Spectral Centroid Magnitude Coefficients (SCMCs)}
Spectral centroid magnitude contains speech information similar to magnitude at the formant frequencies \cite{SpectralMagnitude}. The spectral centroid magnitude coefficients (SCMCs) are computed as follows. First, spectral centroid magnitude (SCM) for the $i$th subband of speech frame is computed as:
\begin{equation}
\label{spectral_centroid_magnitude}
\mathrm{SCM}_i = \frac{\sum_{k=0}^{K/2} f[k]\left| X[k]\right| w_i[k]}{\sum_{k=0}^{K/2}f[k]w_i[k]},
\end{equation}
where $f[k]$ is the normalized frequency ($0\leq f[k]\leq 1$) and $w_i[k]$ is a window function in the frequency domain (here rectangular window is used) for computing the centroid of the $i$-th subband. In the next step, the logarithm of SCM values are computed and converted into feature coefficients (SCMCs) by using DCT. This subband feature outperformed other related features in our preliminary comparison \cite{FeatureComparison}.

\subsection{Constant Q Cepstral Coefficients (CQCCs)}
CQCC is another magnitude-based feature propsed very recently to spoofing detection~\cite{Todisco2016Odyssey}. It was reported to achieve the lowest EERs for known and unknown attacks on the ASVspoof 2015 corpus. CQCC uses a wavelet-like, perceptually motivated time-frequency analysis known as the \emph{constant Q transform} (CQT)~\cite{brown1991calculation}. In contrast to the fixed time-frequency resolution of the short-term Fourier transform, CQT provides a higher frequency resolution for the lower frequencies and a higher temporal resolution for the higher frequencies. In order to compute the cepstrum, the CQT-based power spectrum is first uniformly sampled in linear frequency scale. Finally, CQCCs are computed by performing DCT. In this work, we have used the implementation of CQCC made publicly available by EURECOM\footnote{\url{http://audio.eurecom.fr/software/CQCC_v1.0.zip}}. The default values of the control parameters were used in our experiments.

\subsection{Mean Hilbert Envelope Coefficients (MHECs)}
Gammatone filterbank based features are sometimes used in speech and speaker recognition especially under mismatched and reverberated speech conditions \cite{Sadjadi2015_MHEC, Yin2011_Gammatone, Mitra2014_Gammatone}. In general, the speech signal is first processed by a bank of Gammatone filters that are equally spaced on the equivalent rectangular bandwidth (ERB) scale between 100 and 8000 Hz (assuming the speech signal is sampled at 16 kHz). In this study, we used the Gammatone filterbank implementation provided by Auditory toolbox \cite{AuditoryToolbox}.  

Mean Hilbert envelope coefficients (MHECs) were recently proposed for noise robust speech, speaker, and language recognition \cite{Sadjadi2015_MHEC, Sadjadi2012_MHEC_ASR}. It uses the output of each filter in the filterbank. Calculation of MHEC features is performed through the following steps:
\begin{enumerate}
  \item First, the speech signal is passed through a Gammatone filterbank consisting of 32 filters and for each Gammatone filter output, the temporal envelope, the squared magnitude of the analytical signal is obtained using the Hilbert transform.

  \item The envelope is smoothed by applying a low pass filter with cut-off frequency of $f_c=20$ Hz.

  \item Short-term energy is computed from each smoothed envelope by framing and windowing.

  \item MHECs are computed from the energies using logarithmic compression followed by DCT.
\end{enumerate}

\subsection{Relative-Phase Shift (RPS) Features}
The relative phase shift (RPS) features \cite{DeLeonICASSP2011,DeLeonTASL2012-RPS,RPSTIFS-2015} are based on harmonic modeling of the speech signal. In harmonic modeling, each frame is approximated as the sum of sinusoids in the form:
\begin{equation}
\label{harmonic_model}
x[n] = \sum_k A_k[n]\cos(\phi_k[n]),
\end{equation}
where $A_k[n]$ is the amplitude and
\begin{equation}
\label{instantaneous_phase}
\phi_k[n] = 2\pi k F_0n+\theta_k
\end{equation}
is the instantaneous phase of the $k$th harmonic. $F_0$ is the fundamental frequency and $\theta_k$ is the initial phase of the $k$th harmonic. The instantaneous phase depends on the time instant $n$ and harmonic, $k$, whereas the initial phase, $\theta_k$, is independent of the time instant. The RPS value is the \emph{phase shift} of the $k$th harmonic component with respect to fundamental frequency \cite{DeLeonICASSP2011,DeLeonTASL2012-RPS,RPSTIFS-2015}. It is calculated by solving for $\theta_k$ by equating the time instants $n_i$ in~(\ref{instantaneous_phase}) between the $k$th harmonic and the reference fundamental frequency, assuming $\theta_1=0$:

\begin{equation}
\label{RPS}
\theta_k = \phi_k[n_i] - k\phi_1[n_i],
\end{equation}

We used COVAREP tool \cite{COVAREP} to compute the RPS values. COVAREP tool uses 100 ms frames with 10 ms frame shift for computing the $F_0$. The RPS features are computed from the RPS values by performing phase unwrapping and then differentiation followed by Mel-scale integration and DCT as in \cite{DeLeonICASSP2011, DeLeonTASL2012-RPS}. Similar to other front-end configurations, the $0$th coefficient is included.

\subsection{Modified Group Delay Function}
Group delay function representing phase information shows spurious high amplitude spikes at zeros of short-term magnitude spectrum due to excitation sources. Modified group delay function (MGDF) \cite{Murthy2003ICASSP} is formulated by suppressing zeros of the magnitude spectrum. It is defined as,

\begin{equation}\label{Equation:MODGDF}
\tau(k)=\mathrm{sgn} \times \left| \frac{[X_{\mathrm{R}}(k) Y_{\mathrm{R}}(k) + X_{\mathrm{I}}(k) Y_{\mathrm{I}}(k)]}{H(k)^{2\gamma}}\right|^{\alpha}
\end{equation}

where $\mathrm{sgn}$ is the sign of ${X_{\mathrm{R}}(k) Y_{\mathrm{R}}(k) + X_{\mathrm{I}}(k) Y_{\mathrm{I}}(k)}$, $X_{\mathrm{R}}(k)$ and $X_{\mathrm{I}}(k)$ represent real and imaginary part of DFT for a speech frame $x(n)$ and $Y_{\mathrm{R}}(k)$ and $Y_{\mathrm{I}}(k)$ represent the real and the imaginary parts of DFT for $nx(n)$. $H(k)$ is the speech spectrum after cepstral smoothing, while $\alpha$ and $\gamma$ are two control parameters. Cepstral like features are computed from MGDF using DCT. This feature was used for synthetic speech detection in~\cite{ZhizhengInterspeech2012}. In the experiments, the parameters $\alpha$ and $\gamma$ are set to $0.3$ and $0.1$, respectively.

\subsection{Cosine Phase (CosPhase) Features}
The phase spectrum computed using short-time Fourier transform can be used for speech feature extraction. Since the phase spectrum calculated directly from the complex STFT parameters is discontinuous with respect to frequency, we first unwrap the phase spectrum. The cosine function is then applied to the unwrapped phase spectrum to normalize the range in [-1.0, 1.0]. Then discrete cosine transform (DCT) is applied to the cosine normalized phase spectrum. This feature is called as CosPhase and used in spoofing detection \cite{ZhizhengInterspeech2012}.

\section{Experimental Setup}

\subsection{Database}
\label{database_section}
The experiments are conducted on the ASVspoof 2015 database \cite{ASVspoofOverview} which consists of speech data with no channel or background noise collected from 106 speakers (45 male and 61 female) and three subsets with non-overlapping speakers:
\begin{itemize}
\item \textbf{Training} subset is used to train genuine and spoofed classes for spoofing detection. It contains natural and five different types of spoofed speech: three are generated using voice conversion and the rest using speech synthesis. Voice conversion algorithms are (i) frame-selection (S1), (ii) spectral slope shifting (S2) and (iii) Festvox (S5) system\footnote{\url{http://www.festvox.org}} whereas the speech synthesis spoofs are based on hidden Markov model-based methods (S3 and S4).
\item \textbf{Development} subset is used to optimize spoofing detectors. It contains the same five spoofing methods (S1-S5) as the training subset.
\item \textbf{Evaluation} subset is used for evaluating the final performance of the system. It contains five ``known'' algorithms seen in the training and development subsets (S1-S5) as well as five ``unknown'' algorithms (S6-S10).
\end{itemize}
Table~\ref{table:database} summarizes speaker and utterance information for each subset.

\begin{table}[ht!]
 \centering
 \begin{small}
 \caption{Statistics of the ASVspoof 2015 database, used in the experiments \cite{ASVspoofOverview}.}
 \label{table:database}
\medskip
 \begin{tabular}{l||cc||cc}
  \noalign{\hrule height 0.75pt}
 \multirow{2}{*}{Subset} & \multicolumn{2}{c||}{Number of speakers} & \multicolumn{2}{c}{Number of utterances}\\
 & \emph{Male} & \emph{Female} & \emph{Natural} & \emph{Synthetic} \\   \noalign{\hrule height 0.75pt}
 Training & 10 & 15 & 3750 & 12625 \\
 Development & 15 & 20 & 3497 & 49875 \\
 Evaluation & 20 & 26 & 9404 & 184000 \\
 \noalign{\hrule height 0.75pt}
 \end{tabular}
 \end{small}
\end{table}

\begin{figure}
\begin{centering}
\includegraphics[scale=0.3]{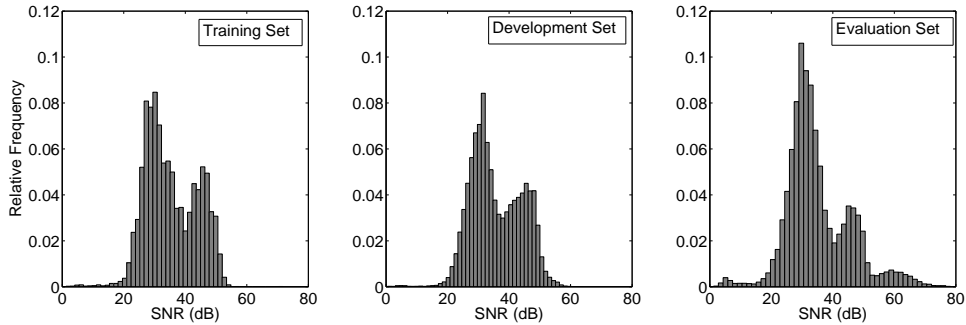}
\caption{Distributions of estimated SNR levels for each subset of ASVspoof 2015 dataset.}
\label{ASVSpoof_SNR}
\end{centering}
\end{figure}

To analyze the original ASVspoof 2015 data regarding noise level and to show the quality of recordings in the database before interpreting the results, we computed the SNR level of recordings. Fig.~\ref{ASVSpoof_SNR} shows the histograms of estimated SNR levels\footnote{SNREval Toolkit from \url{http://labrosa.ee.columbia.edu/projects/snreval/} is used to estimate the SNR levels.} for each subset of the original ASVspoof 2015 dataset. All the speech files from the training set are used to plot the histogram for this subset, whereas randomly selected $30000$ speech signals are used to generate histograms for Evaluation and Development subsets. A vast majority of the signals have a relatively high SNR exceeding $20$ dB. The evaluation subset contains also signals with very high SNRs (approximately $8\%$ of $30000$ files have SNR $>50$ dB).

We use \emph{Filtering and Noise Adding Tool} (FaNT)\footnote{\url{http://dnt.kr.hsnr.de/}} to corrupt the original ASVspoof 2015 signals with noise for introducing controlled degradation. FaNT is an open-source tool which follows the ITU recommendations for noise adding and filtering. To be more precise, it uses psychoacoustic speech level computation based on the ITU recommendation P.56 (\emph{objective measurement of active speech level}). We digitally add white, babble and car noises from NOISEX-92 database \cite{Noisex92}. For each noise type we consider 3 SNR levels: $0$, $10$ and $20$ dB. The reasons for selecting these types of noise are the following: (i) White noise has a flat spectral density and it masks all the frequency components uniformly. Although it rarely represents a real-case scenario, it is a commonly used control noise in studying robust speech processing methods. (ii) Babble noise is one of the most difficult noise types in speech applications containing a mixture of multiple speakers~--- a situation that occurs on a daily basis in any crowded place \cite{babble_noise}. (iii) Car noise is another noise type that may frequently occur in our daily life such as making a phone call while driving.

In the experiments, we consider \textbf{noise mismatched} condition by training the natural and synthetic speech models using the original clean training files, but test them on noisy files. The reason for this choice is practicality: in a real-world deployment of ASV technology in smartphones or other portable devices, the operation environment of the user would be rarely known precisely.

\subsection{Classifier and Features}
We use 32 coefficients (including $c_0$) and 32 filters in filterbank for every method. This is done to have comparable results for different feature extraction methods. We apply energy-based voice activity detection (VAD) \cite[p. 24]{KinnunenOverview} on clean data to get speech/non-speech labels. Using clean VAD labels allows us to focus merely on the effect of noise on synthetic speech detection rather than mixed effects of VAD and feature set. These labels are used to discard non-speech frames for both clean and noisy speech.

For GMM-based classification, we use two models to represent natural and synthetic speech classes (see details in Section~\ref{sec:spoof_detect}).
GMM for each class has 512 components and is trained using 5 EM iterations (the performance differences for larger number of components were negligible in our initial experiments).

For i-vector based classification, we train a gender-independent universal background model (UBM) consisting of 512 Gaussians using $9000$ utterances from $150$ male and $150$ female speakers from the WSJ0 \& 1 corpora \cite{WSJ}. To train the $\mathrm{\mathbf{T}}$-matrix, we select $8945$ utterances from 178 male and 177 female speakers from the WSJ0 \& 1 databases\footnote{Usually 283 speakers from WSJ0 \& 1 databases are used in most studies which is the official training set of the corpora. In our experiments, we included test sets of WSJ0 \& 1 corpora in addition to training set which yields a total of 177 male and 178 female speakers.} and run EM-algorithm for 5 iterations. The extracted 600 dimensional i-vectors are further processed by applying \emph{within-class covariance normalization} (WCCN) \cite{WCCN}, followed by projection to the unit sphere \cite{LengthNorm}. The logic behind WCCN is not to normalize within-speaker variation \cite{NajimIvector}, like it is done for speaker recognition, but to normalize within-class (natural or synthetic) variation. To this end, we separate the training data into natural and synthetic classes and use them to compute WCCN transformation matrix $\mathbf{B}$ \cite[p. 791]{NajimIvector}. PLDA model trained on original (clean) data is used in noisy spoofing detection experiments.

\subsection{Combined Countermeasures via Score Fusion}

Given the wide diversity and varied difficulty of existing and future spoofing attacks, it might be difficult to come up with a single feature set to detect all possible attacks. As an example, phase-related features might be suited to detect attacks whose vocoder discards natural phase information while other methods may possess superior noise robustness. This motivates exploration towards countermeasures that includes a bank of different front-ends, some being potentially specialized to detect particular types of attacks. To this end, here we consider two score level fusion strategies to maximally benefit from the complementarity of our features: 1) \textbf{Fusion 1: Score averaging}~--- a simple technique, which does not require any training, 2) \textbf{Fusion 2: weighted sum}, where fusion weights and a bias term are estimated using logistic regression \cite{Brummer2007_fusion}. We use the development data to train the parameters for each noise type and SNR level.

\subsection{Performance Measure}
Following the evaluation plan of ASVspoof 2015, equal error rate (EER) is used as the objective performance criterion in the experiments. EER corresponds to the threshold at which false acceptance ($P_{\mathrm{fa}}$) and miss rate ($P_{\mathrm{miss}}$) are equal. $P_{\mathrm{fa}}$ is the ratio of number of spoof trials detected as genuine speech to the total number of spoof trials and $P_{\mathrm{miss}}$ is the ratio of number of genuine trials detected as spoofed to the total number of genuine trials. The EERs reported in this work were computed using the bosaris toolkit\footnote{\url{https://sites.google.com/site/bosaristoolkit/}} which computes the EER on receiver operating characteristics (ROC) convex hull (ROCCH) that is an interpolated version of standard ROC.

\section{Results}
\label{Results}
We conduct the experiments separately on the development and evaluation parts of ASVspoof 2015. The development part is first used for optimizing the system parameters and configurations. The feature extraction method that yield the lowest EERs is then selected for further experiments on the evaluation part.

\begin{figure}
\begin{centering}
\includegraphics[scale=0.35]{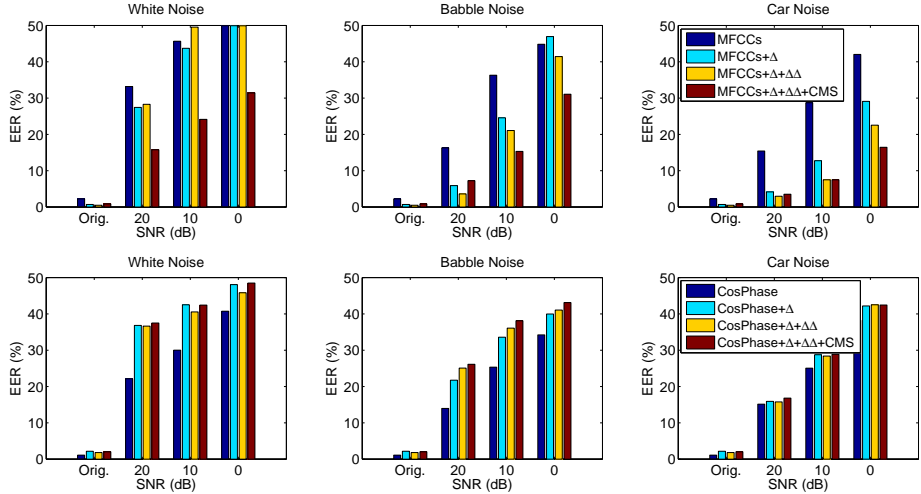}
\caption{Effects of $\Delta$ and $\Delta\Delta$ MFCC features and Cepstral Mean Subtraction on synthetic speech detection. First row, MFCC features. Second row, CosPhase Features.}
\label{MFCCPostProcessingGraph}
\end{centering}
\end{figure}
\subsection{Effect of Feature Post-Processing}
In our first experiment on the development set, we study the effect of feature post-processing. Specifically, we study the appending $\Delta$ and $\Delta\Delta$ features and cepstral mean subtraction (CMS). The results on MFCC and CosPhase features are shown in Fig.~\ref{MFCCPostProcessingGraph}. Here, MFCC and CosPhase features are selected as representatives of magnitude and phase-based features, respectively. The upper row corresponds to the MFCC and the lower row to the CosPhase features. For the original (clean) case, $2.24\%$ EER is obtained using only the base MFCCs. Appending $\Delta$ and $\Delta\Delta$ features to the MFCCs reduces the EER to $0.49\%$. Applying CMS slightly reduces the performance for the clean case ($0.84\%$ EER). For the CosPhase features, in turn, the lowest EER ($1.09\%$) is obtained with the base features on clean data in contrast to MFCCs. Appending $\Delta$ features to the base CosPhase features almost doubles the EER ($2.16\%)$. Appending $\Delta\Delta$ or applying CMS does not help to increase the synthetic speech detection performance with CosPhase features.

For the noisy case, appending the $\Delta$ and $\Delta\Delta$ coefficients considerably improves the accuracy in most cases. For example, we see $78\%$ relative improvement over the base MFCCs for babble noise at $20$ dB SNR (EER $16.29\%\: \rightarrow \:3.61\%$). Similarly, applying CMS on top of the dynamic features improves performance considerably. Whereas, post-processing shows an opposite effect with CosPhase features where the smallest EERs are obtained with the base features independent of the noise type and SNR.

The results in Fig.~\ref{MFCCPostProcessingGraph} are for the MFCC and CosPhase features. The results were similar for the other studied features. Namely, for the magnitude (MFCCs, IMFCCs, SCMC and CQCC) and MHEC features, the best performance is obtained with the full post-processing (included deltas followed by CMS) whereas for the phase-based features the raw features yield the smallest EERs except for MGD. Out from the 10 conditions evaluated (3 SNRs $\times$ 3 noise types plus the clean data), MGD features with deltas and feature normalization yielded the lowest EERs in 6 cases. Thus, in all the remaining experiments, we will adopt the raw RPS and CosPhase features. For all the rest of the features, we include deltas and CMS.

\begin{table}[!t]
\caption{Comparison (EER, \%) of different front-end features in noisy conditions on development set using Gaussian Mixture Model classifier. The results for clean original condition are presented as well as the average results for all noisy sub-conditions.}
\begin{center}
\footnotesize
\begin{tabular}{cc||cccccccc||cc}
\noalign{\hrule height 0.75pt}
Noise & SNR & \multirow{2}{*}{MFCC} & \multirow{2}{*}{IMFCC} & \multirow{2}{*}{SCMC} & \multirow{2}{*}{CQCC} & \multirow{2}{*}{MHEC} & \multirow{2}{*}{RPS} & \multirow{2}{*}{MGD} & \multirow{2}{*}{CosPhase} & \multirow{2}{*}{Fusion1} & \multirow{2}{*}{Fusion2} \\
type & (dB) & & & & & & & & & \\ \noalign{\hrule height 0.75pt}
\multicolumn{2}{c||}{Original} & 0.84 & 0.91 & 0.38 & 0.44 & 3.92 & \textbf{0.15}  & 1.25 & 1.09 & 0.02 & \textbf{0.00} \\ \noalign{\hrule height 0.75pt}
\multirow{3}{*}{White} & 20 & 15.75 & 34.17 & 21.91 & 33.41 & \textbf{12.08} & 37.64 & 28.35 & 22.12 & 12.17 & \textbf{8.45}  \\
& 10 & 24.13 & 44.56 & 32.19 & 38.13 & \textbf{22.2} & 41.37 & 39.23 & 30.02 & 18.84 & \textbf{16.09} \\
& 0 & \textbf{31.42} & 48.86 & 39.86 & 45.55 & 33.37 & 43.61 & 46.45 & 40.73 & 29.42 & \textbf{27.69} \\ \noalign{\hrule height 0.75pt}
\multirow{3}{*}{Babble} & 20 & 7.23 & 5.66 & \textbf{2.71} & 18.07 & 11.06 & 5.26 & 13.77 & 13.97 & 1.89 & \textbf{0.56} \\
& 10 & 15.32 & 15.4 & \textbf{9.36} & 29.49 & 25.58 & 20.04 &  26.26 & 25.33 & 7.72 & \textbf{4.96} \\
& 0 & 31.05 & 37.73 & \textbf{30.09} & 41.60 & 40.87 & 39.90 &  40.12 & 34.22 & 26.58 & \textbf{22.85} \\ \noalign{\hrule height 0.75pt}
\multirow{3}{*}{Car} & 20 & 3.51 & 1.94 & 0.87 & 9.26 & 8.96 & \textbf{0.74} & 9.30 & 15.14 & 0.39 & \textbf{0.03} \\
& 10 & 7.48 & 4.69 & \textbf{2.48} & 18.04 & 19.47 & 5.75 & 15.84 & 25.05 & 2.56 & \textbf{0.67}\\
& 0 & 16.44 & 14.27 & \textbf{8.74} & 29.42 & 33.12 & 24.03 &  29.72 & 38.23 & 11.83 & \textbf{7.12} \\ \noalign{\hrule height 0.75pt}
\multicolumn{2}{c||}{Average} & 16.92 & 23.03 & \textbf{14.85} & 26.34 & 21.06 & 21.84 & 25.02 & 24.58 & 11.14 & \textbf{8.84} \\
\noalign{\hrule height 0.75pt}
\end{tabular}
\end{center}
\label{comp_features_table}
\end{table}

\subsection{Comparison of Features}
The results on development set for different features using GMM are summarized in Table~\ref{comp_features_table}. For the clean (original) case, the RPS features yield the lowest EER. However, under additive noise, especially for white noise and at low SNR levels of car and babble noises, the performance of RPS is relatively poor. This could be because RPS requires estimated $F_0$ values that are difficult to extract reliably from noisy data. For babble and car noises at high SNRs ($20$ dB), RPS yields reasonable accuracy. The SCMC features perform well for the babble and car noises, whereas for white noise, MHEC yields lower EERs. To sum up Table~\ref{comp_features_table}, none of the feature sets is consistently superior to others. In most cases, SCMC outperforms the other features. Out of the three phase features (RPS, MGD and CosPhase), CosPhase features are superior to RPS and MGD under white noise case. However, RPS outperforms MGD and CosPhase for babble and car noises. In general, magnitude features outperform phase-related features independent of noise type and SNR.

Applying score fusion to the eight feature extraction methods considerably improves the accuracy for all cases including the original (clean) condition as Table~\ref{comp_features_table} indicates. Weighted sum technique where the weights of each individual system are estimated with logistic regression (indicated as Fusion2 in Table~\ref{comp_features_table}) yield lower EERs than score averaging fusion (Fusion1). The effect of each individual feature set on the fusion performance has been investigated and it was found that excluding RPS from the fusion (applying score fusion to the six remaining feature sets) dramatically increases the EERs irrespective of noise and SNR. This suggests that RPS consists of complementary information even though it gives poor stand-alone performance compared to other features.

\subsection{Effect of Speech Enhancement}
\begin{figure}[ht] 
\centering
\begin{subfigure}{\includegraphics[width=.3\linewidth] {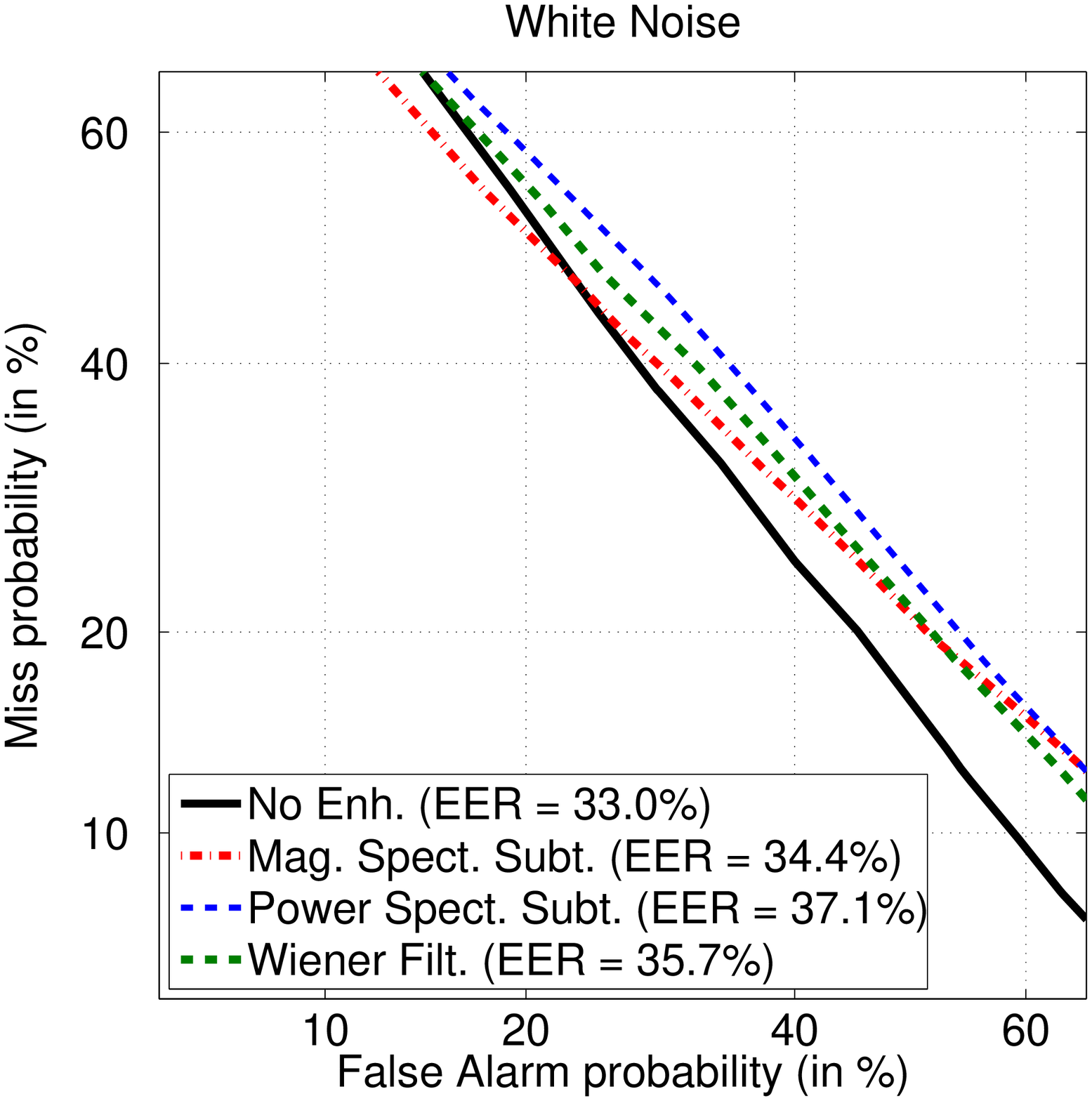}
   \label{fig:subfig1}
 }%
\end{subfigure}\hfill
\begin{subfigure}{\includegraphics[width=.3\linewidth] {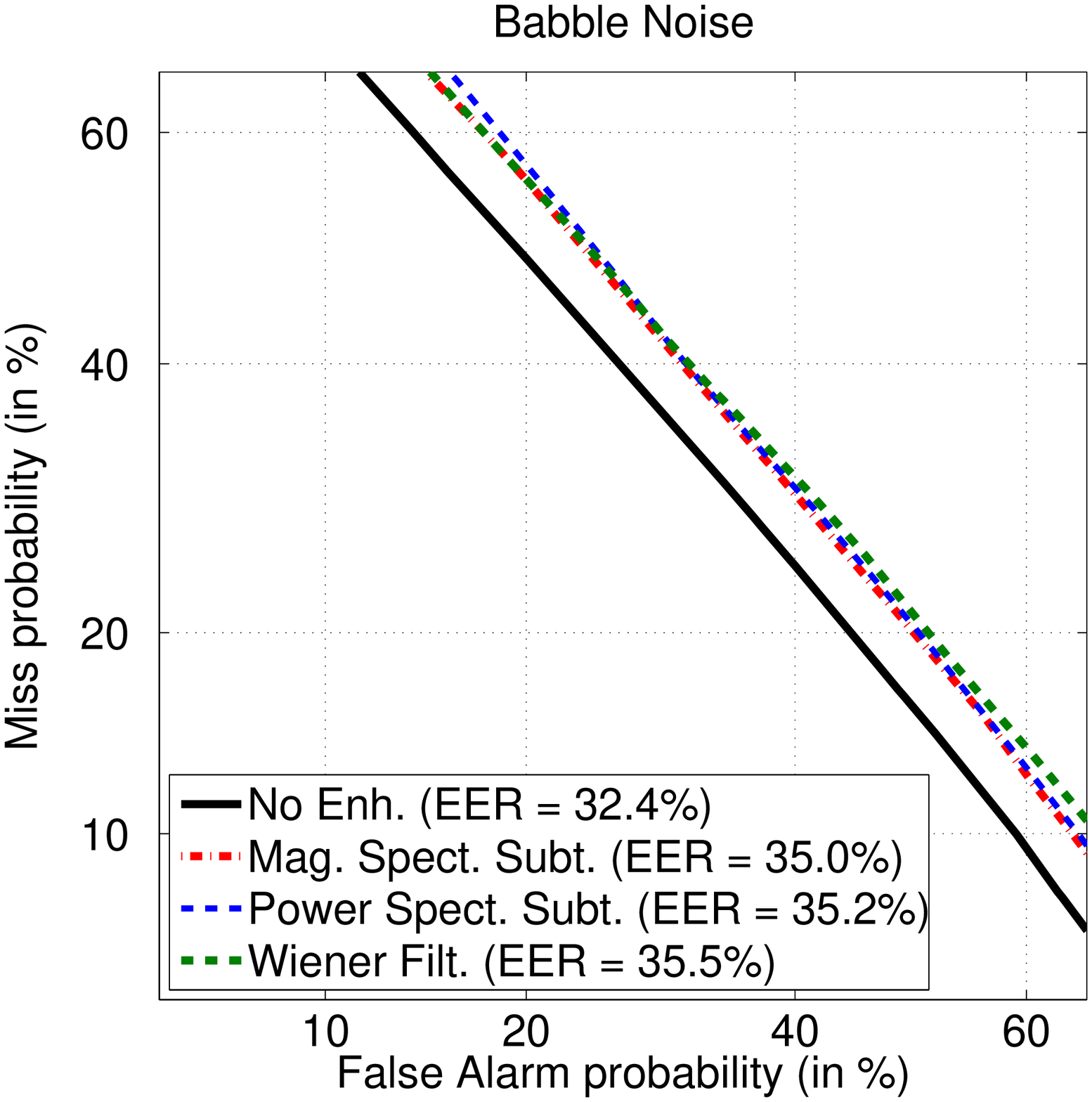}
   \label{fig:subfig2}
 }%
\end{subfigure}\hfill
\begin{subfigure}{\includegraphics[width=.3\linewidth] {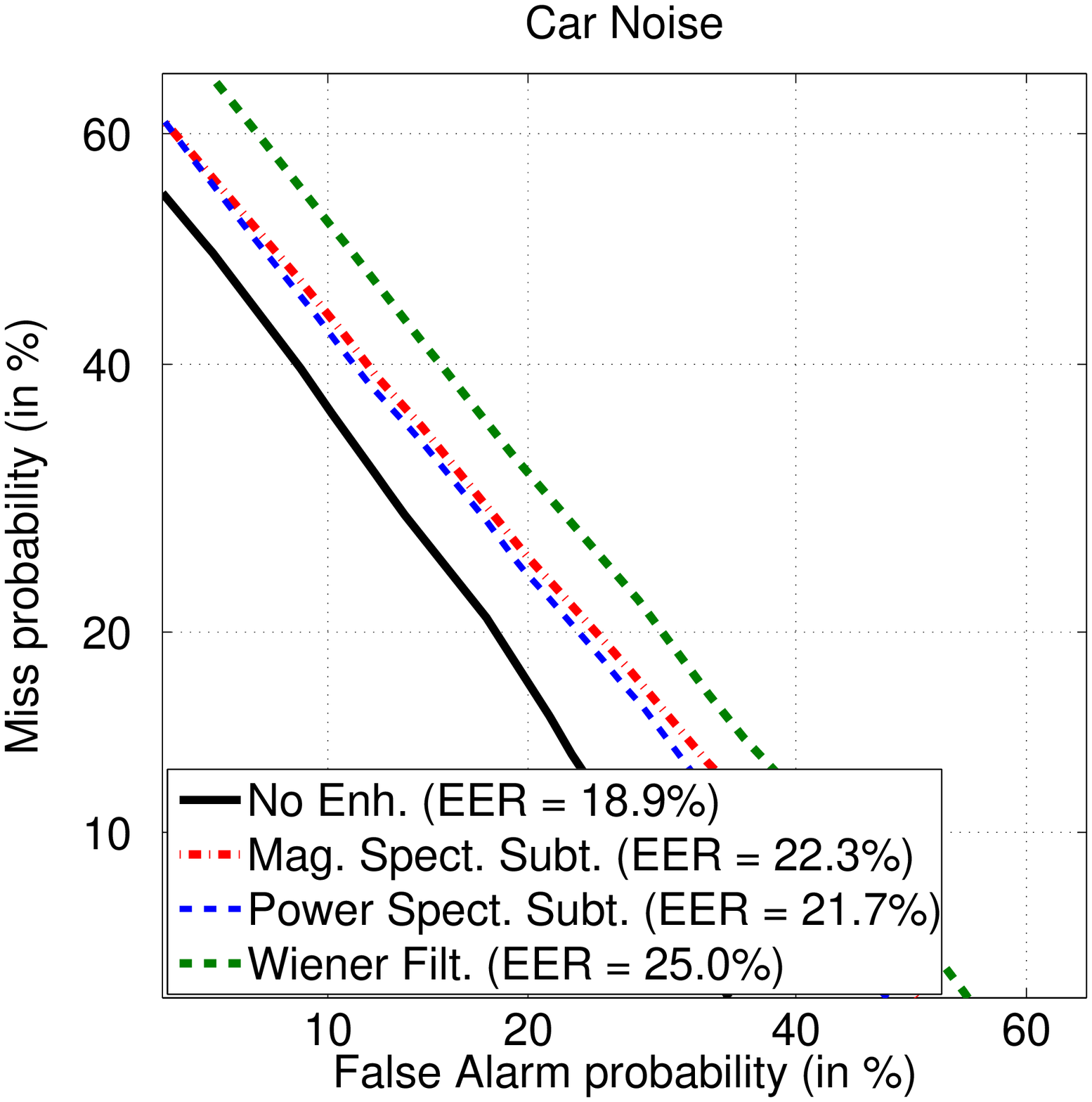}
   \label{fig:subfig3}
 }%
\end{subfigure}\hfill

\medskip
\begin{subfigure}{\includegraphics[width=.3\linewidth] {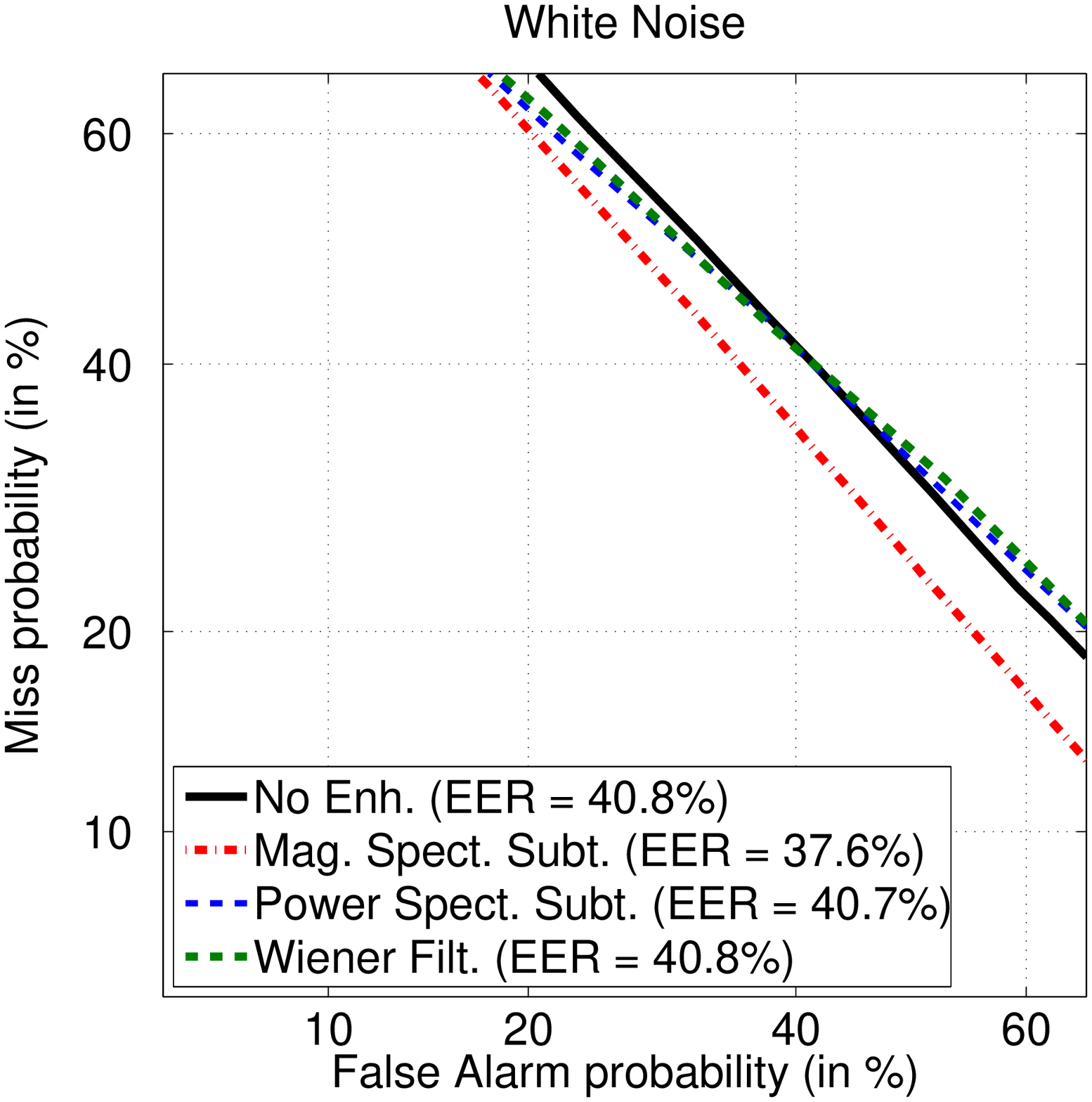}
   \label{fig:subfig4}
 }%
\end{subfigure}\hfill
\begin{subfigure}{\includegraphics[width=.3\linewidth] {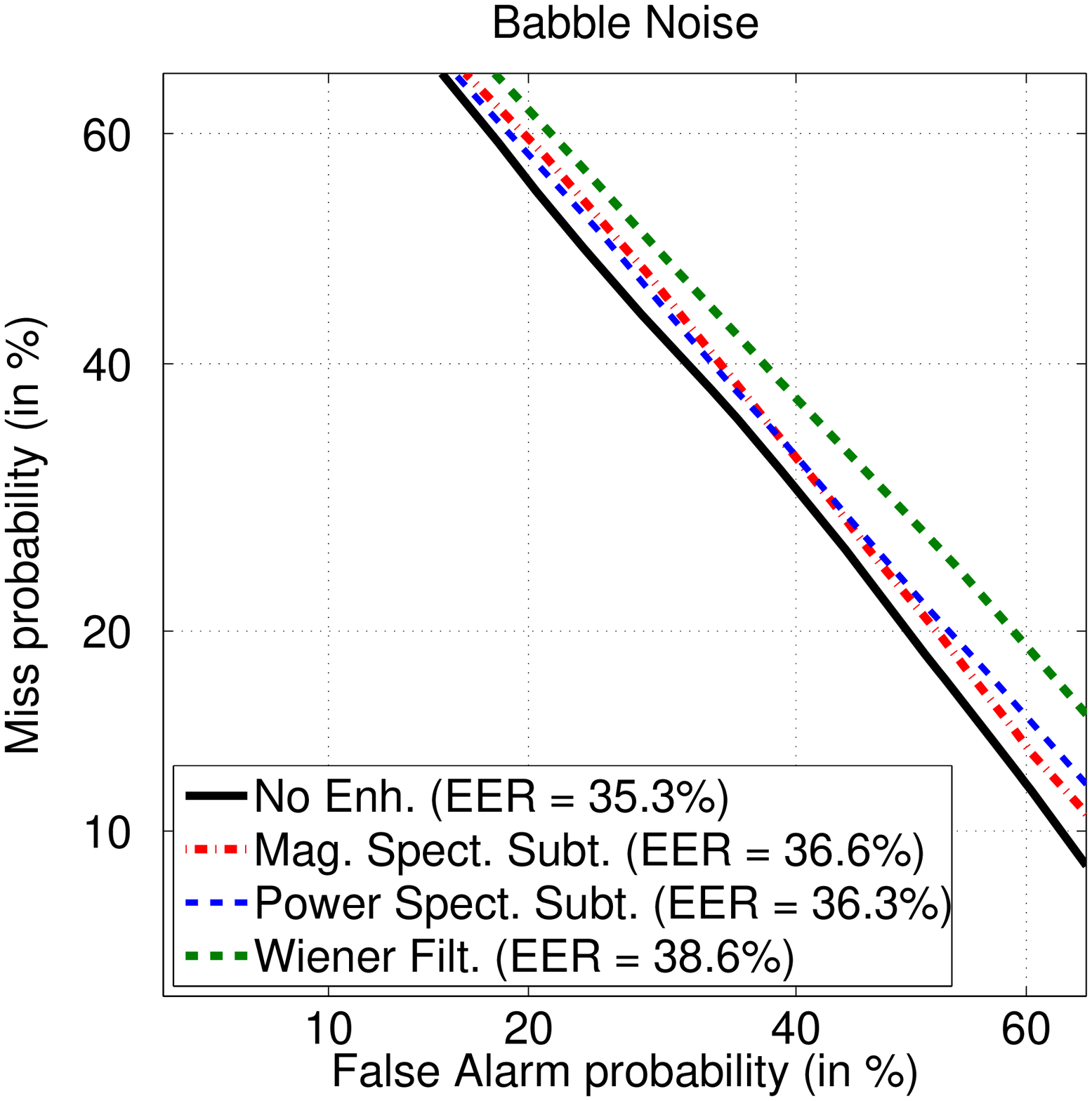}
   \label{fig:subfig5}
 }%
\end{subfigure}\hfill
\begin{subfigure}{\includegraphics[width=.3\linewidth] {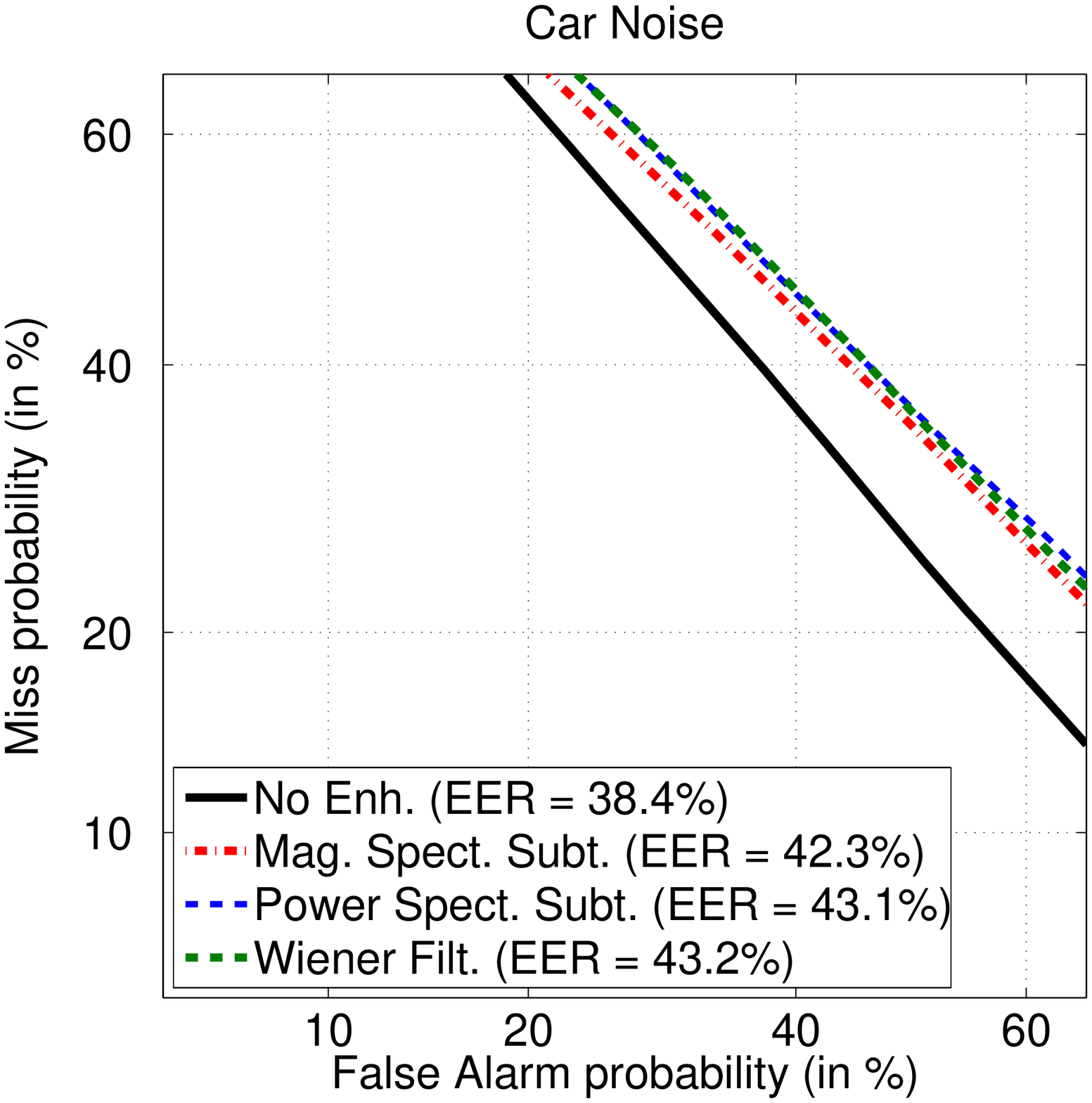}
   \label{fig:subfig6}
 }%
\end{subfigure}
\caption{DET curves for different speech enhancement techniques under additive noise (0 dB). First row, MFCC features. Second row, CosPhase features.}
\label{DETCurvesSpeechEnhancement}
\end{figure}

Next, we study the effect of speech enhancement techniques. To this end, magnitude and power spectral subtraction algorithms \cite{MagSpectralSub,PowerSpectralSub} and Wiener filtering \cite{WienerFiltering} approaches are adopted. Detection error trade-off (DET) curves for different speech enhancement methods for each noise type, at $0$ dB SNR and using MFCC features with deltas and CMS as well as CosPhase features, are shown in Fig.~\ref{DETCurvesSpeechEnhancement}. Here, the DET curves are generated by pooling the scores of all the individual attacks\footnote{Although in ASVspoof 2015 the evaluation metric is averaged EER over different attacks, producing a single DET curve that would coincidence with this operating point is not obvious. Thus, here the scores are pooled to generate the DET plot and to compute the corresponding EERs in Fig.~\ref{DETCurvesSpeechEnhancement} legends.}. Fig.~\ref{DETCurvesSpeechEnhancement} indicates that the attempted speech enhancement techniques do not yield any performance gains for MFCC features. For CosPhase features, magnitude spectral subtraction slightly improves the performance for white noise whereas for babble and car noises speech enhancement methods do not improve the performance. These three methods were applied to SCMC features as well in order to analyse the effect of speech enhancement on different features and to check whether the observations can be generalized and the similar results have been obtained. Apart from these three popular methods, other methods including minimum mean square error (MMSE), logarithmic MMSE (logMMSE) and iterative Wiener filtering techniques (as available in the Appendix of \cite{LoizouSpeechEnhancement}) were studied, without success. The reduction on the performance after speech enhancement might be because speech enhancement introduces musical noise and other processing artifacts that mask the synthesis or conversion artifacts.

A recent independent study ~\cite{YU2016} confirms the ineffectiveness of traditional unsupervised speech enhancement techniques for spoofing detection in noisy condition. Currently, similar to most speech processing tasks, the use of deep neural network (DNN) based techniques is extensively studied on speech enhancement ~\cite{Xu2015,Xu2014,Han2015} and could be an interesting approach. However, as DNNs require large amounts of additional training data from different noisy conditions for supervised training, they are not addressed in this study that focuses on DSP-based unsupervised speech enhancement techniques. Further, achieving performance improvement in unseen noisy condition appears challenging even with DNN-based speech enhancement methods~\cite{Sun2016}.

\begin{table}[!h]
\caption{Comparison (EER, \%) of different front-end features in noisy conditions on development set using Cosine/Probabilistic Linear Discriminant Analysis i-vector classifiers. The results for the clean (original) condition are presented as well as the average results for all noisy sub-conditions. The lower half of the table presents a difference between the corresponding EERs for PLDA and cosine scoring. Blue cells indicate conditions where PLDA scoring is advantageous to cosine scoring, whereas red cells indicate the opposite.}
\begin{center}
\footnotesize
\begin{tabular}{cccccccccc||cc}
\multicolumn{11}{c}{\textbf{Cosine Scoring}} \\
& SNR (dB) & MFCC & IMFCC & SCMC & CQCC & MHEC & RPS & MGD & CosPhase & Fusion1 & Fusion2 \\ \noalign{\hrule height 0.75pt}
& Original & 5.12 & \textbf{3.24} & 5.30 & 0.26 & 12.31  & 5.18 & 8.40 & 11.80 & 0.01 & \textbf{0.00} \\ \noalign{\hrule height 0.75pt}
\multirow{3}{*}{\rotatebox[origin=c]{90}{White}} & 20 & 26.48 & 45.51 & 39.97 & 41.55 & \textbf{26.05}  & 39.97 & 39.34 & 32.61 & 20.37 & \textbf{17.27} \\
& 10 & 36.35 & 47.60 & 44.15 & 44.76 & \textbf{30.71} & 45.24 & 45.70 & 35.98 & 31.4 & \textbf{26.26} \\
& 0 & 43.47 & 48.26 & 46.68 & 48.27 & \textbf{39.20} & 47.60 & 48.04 & 47.98 & 41.55 & \textbf{37.53} \\ \noalign{\hrule height 0.75pt}
\multirow{3}{*}{\rotatebox[origin=c]{90}{Babble}} & 20 & 20.94 & 28.07 & 24.44 & 27.63 & 25.58 & \textbf{19.10} & 25.12 & 33.11 & 6.15 & \textbf{5.75} \\
& 10 & 33.59  & 40.54  & 34.97 & 39.21 & 33.54  & \textbf{31.03} & 36.37 & 41.23 & 18.71 & \textbf{18.13} \\
& 0 & 45.71  & 48.15  & 45.02 & 46.20 & 43.65  & 43.73 & 45.59 & \textbf{43.47} & 38.88 & \textbf{37.57} \\ \noalign{\hrule height 0.75pt}
\multirow{3}{*}{\rotatebox[origin=c]{90}{Car}} & 20 & 24.00  & 13.56 & 14.34 & 13.46 & 22.53  & \textbf{11.88} & 21.42 & 33.84 & 2.04 & \textbf{1.86} \\
& 10 & 33.67  & 26.28 & 22.61 & 25.53 & 27.76  & \textbf{22.14} & 29.78 & 43.91 & 8.02 & \textbf{8.01} \\
& 0 & 39.62  & 40.39  & \textbf{33.12} & 37.84 & 34.76  & 38.34 & 38.82 & 48.45 & 23.18 & \textbf{22.24} \\ \noalign{\hrule height 0.75pt}
\multicolumn{2}{c}{Average} & 30.89 & 34.16  & 31.06 & 32.47 & 29.60  & 30.42 & 33.85 & 36.75 & 19.03 & \textbf{17.46} \\ \noalign{\hrule height 0.75pt} \\

\multicolumn{11}{c}{\textbf{PLDA - Cosine}} \\
& SNR (dB) & MFCC & IMFCC & SCMC & CQCC & MHEC & RPS & MGD & CosPhase & Fusion1 & Fusion2 \\ \noalign{\hrule height 0.75pt}
& Original & \cellcolor{blue!25} -0.09 &\cellcolor{red!25} 0.81 &\cellcolor{red!25} 0.73 & \cellcolor{blue!25} -0.01 & \cellcolor{red!25} 0.39 &\cellcolor{blue!25}-0.15 &\cellcolor{red!25} 0.08 &\cellcolor{blue!25}-7.26 &\cellcolor{red!25} 0.01 &\cellcolor{red!25} 0.00  \\ \noalign{\hrule height 0.75pt}
\multirow{3}{*}{\rotatebox[origin=c]{90}{White}} & 20 &\cellcolor{blue!25}-1.16 &\cellcolor{blue!25}-0.87 &\cellcolor{blue!25}-0.93 & \cellcolor{red!25} 2.21 & \cellcolor{red!25} 2.18 &\cellcolor{red!25} 3.12 &\cellcolor{red!25} 0.06 &\cellcolor{red!25} 3.12 &\cellcolor{red!25} 1.69 &\cellcolor{red!25} 1.88 \\
& 10 &\cellcolor{blue!25} -1.48 &\cellcolor{red!25} 0.14 &\cellcolor{blue!25} -0.04 & \cellcolor{red!25} 1.25 & \cellcolor{red!25} 2.49 &\cellcolor{red!25} 1.53 & \cellcolor{blue!25}-0.27 & \cellcolor{red!25}0.7 &\cellcolor{blue!25} -1.40 &\cellcolor{red!25} 2.44 \\
& 0 &\cellcolor{red!25} 0.2 &\cellcolor{red!25} 0.32 &\cellcolor{red!25} 0.27 & \cellcolor{red!25} 0.3 & \cellcolor{red!25} 1.29 &\cellcolor{blue!25} -0.16 &\cellcolor{blue!25} -0.15 & \cellcolor{blue!25} -5.69 &\cellcolor{blue!25} -1.99 &\cellcolor{red!25} 0.65 \\ \noalign{\hrule height 0.75pt}
\multirow{3}{*}{\rotatebox[origin=c]{90}{Babble}} & 20 &\cellcolor{blue!25} -0.29 &\cellcolor{blue!25} -0.04 &\cellcolor{red!25} 0.04 & \cellcolor{red!25} 0.84  & \cellcolor{red!25} 0.49 &\cellcolor{red!25} 4.05 &\cellcolor{red!25} 0.7 &\cellcolor{red!25} 2.81 &\cellcolor{red!25} 2.87 &\cellcolor{red!25} 1.35 \\
& 10 &\cellcolor{blue!25} -0.46 &\cellcolor{blue!25} -0.92 &\cellcolor{blue!25} -0.42 &\cellcolor{red!25} 0.94  & \cellcolor{red!25}0.12 & \cellcolor{red!25}3.99 &\cellcolor{blue!25} -0.13 &\cellcolor{red!25} 1.18 &\cellcolor{red!25} 4.83 &\cellcolor{red!25} 1.52\\
& 0 &\cellcolor{red!25} 0.11 &\cellcolor{red!25} 0.08 &\cellcolor{blue!25} -0.36 & \cellcolor{blue!25} -0.19 & \cellcolor{blue!25} -0.12 & \cellcolor{red!25} 1.69 &\cellcolor{blue!25} -0.31 &\cellcolor{red!25} 0.77 &\cellcolor{red!25} 0.58 &\cellcolor{red!25} 3.01 \\ \noalign{\hrule height 0.75pt}
\multirow{3}{*}{\rotatebox[origin=c]{90}{Car}} & 20 &\cellcolor{red!25} 0.35 &\cellcolor{red!25} 0.97 &\cellcolor{red!25} 0.55 &\cellcolor{red!25} 0.61 &\cellcolor{red!25} 0.24 &\cellcolor{red!25} 1.91 &\cellcolor{red!25} 12.42 &\cellcolor{red!25} 2.17 &\cellcolor{red!25} 1.59 &\cellcolor{red!25} 0.72 \\
& 10 &\cellcolor{blue!25} -1.17 &\cellcolor{blue!25} -0.23 &\cellcolor{red!25} 0.53 & \cellcolor{red!25} 1.34 & \cellcolor{red!25} 0.51 &\cellcolor{red!25} 3.97 & \cellcolor{red!25}0.28 & \cellcolor{red!25}0.54 &\cellcolor{red!25} 3.79 &\cellcolor{red!25} 0.65 \\
& 0 &\cellcolor{blue!25} -1.61 &\cellcolor{blue!25} -1.32 &\cellcolor{blue!25} -0.61 & \cellcolor{red!25} 0.22 &\cellcolor{red!25} 0.82 &\cellcolor{red!25} 3.85 &\cellcolor{blue!25} -0.38 &\cellcolor{blue!25} -0.34 &\cellcolor{red!25} 5.76 &\cellcolor{red!25} 0.56 \\ \noalign{\hrule height 0.75pt}
\multicolumn{2}{c}{Average} &\cellcolor{blue!25} -0.56 &\cellcolor{blue!25} -0.11 &\cellcolor{blue!25} -0.03 & \cellcolor{red!25} 0.75 & \cellcolor{red!25} 0.85 &\cellcolor{red!25} 2.38 &\cellcolor{red!25} 0.11 &\cellcolor{red!25} 0.28 &\cellcolor{red!25} 1.77 &\cellcolor{red!25} 1.27 \\
\end{tabular}
\end{center}
\label{comp_features_table_ivector}
\end{table}

\subsection{i-vector Countermeasures from Different Features}
Up to this point, we have utilized the computationally light GMM classifier to study different feature configurations. In our last experiments with the development set, we study an i-vector based countermeasure. To this end, i-vector extractors are trained from scratch for all the seven acoustic feature sets. The results are provided in Table~\ref{comp_features_table_ivector} for both cosine and PLDA scoring. For clean (original) case, the recently proposed CQCC features yield the smallest EER among the eight methods. While the performance of CQCC features with i-vector back-end is superior to GMM classifier on clean data, for the remaining seven feature extraction methods, GMM back-end outperforms the i-vector back-end. For additive noise cases, i-vector is inferior to GMM independent of the noise type and feature extraction method. Similar results for GMM and i-vector techniques were found in our recent comparative study of classifiers for synthetic speech detection \cite{ClassifierComparison}. This could be because of the short duration of recordings (approximately 3 seconds) that ASVspoof 2015 consists of. Similar observation for i-vector performance on short utterances were found in \cite{ShortData} where GMM and i-vector systems were compared for speaker verification task using short data and it was found GMM recognizer outperforms i-vector system.

Similar to GMM experiments under additive noise (Table~\ref{comp_features_table}), none of the features are systematically superior to others. The features that yield the lowest EERs are different for each noise type and SNR level. MHEC yields the highest performance for white noise whereas, for the babble and car noises, RPS is superior to other features at high SNRs ($20$ and $10$ dB). Concerning the two i-vector back-end variants, PLDA does not bring substantial improvements in comparison to cosine scoring. The most considerable performance improvement with PLDA is obtained with CosPhase features using original (clean) data (EER reduced from $11.80\%$ to $4.54\%$ with PLDA). Similar to the results with GMM classifier, CosPhase features outperform the other phase features (RPS and MGD) under white noise. However, for the babble and car noises, RPS outperforms other phase features. The performance of MGD features, in turn, lies between RPS and CosPhase. In the next experiments on Evaluation set, MFCC and SCMC features as two magnitude and RPS and MGD features as phase based features using GMM and i-vector techniques will be considered.

\subsection{Results on Evaluation Set}
In the experiments with the evaluation portion of ASVspoof 2015, we first study the performance of each individual attack using clean data with two magnitude (MFCC and SCMC) and two phase (RPS and MGD) based features.
The EERs obtained with GMM and i-vector techniques for the individual attacks are summarized in Table~\ref{evaluation_set_individual_clean_results}.
Similar to observations found on the development set, GMM outperforms both i-vector scoring variants independent of the attack type and the features.

\begin{table}[!h]
\caption{
Comparison (EER, $\%$) of Gaussian Mixture Model classifier and two i-vector based classifiers: Cosine scoring and Probabilistic Linear Discriminant Analysis.
We consider individual attacks on clean evaluation set using selected two magnitude (MFCCs and SCMC) and two phase (RPS and MGD) based features.}
\begin{center}
\footnotesize
\begin{tabular}{cl||ccccc||ccccc||c}
\noalign{\hrule height 0.75pt}
\multirow{2}{*}{Features} & \multirow{2}{*}{Classifier} & \multicolumn{5}{c||}{\textbf{Known Attacks}} & \multicolumn{5}{c||}{\textbf{Unknown Attacks}} & Avg. \\
& & S1 & S2 & S3 & S4 & S5 &  S6 & S7 & S8 & S9 & S10 & (S6-S9) \\ \noalign{\hrule height 0.75pt}
\multirow{3}{*}{\rotatebox[origin=c]{90}{MFCC}} & GMM & \textbf{0.00} & 3.54 & \textbf{0.00} & \textbf{0.00} & 0.70 & 1.10 & 0.80 & 0.53 & 0.11 & 27.34 & 0.63 \\
& Cosine & 2.89 & 9.26 & 2.67 &2.66 & 6.01 & 8.07 &3.64 & 5.03 & 3.07 & 46.49 & 4.95 \\
& PLDA & 3.26 & 9.67 & 2.16 & 2.39 & 5.84 & 8.23 &3.55 & 6.97& 3.29 & 47.11 & 5.51 \\ \hline

\multirow{3}{*}{\rotatebox[origin=c]{90}{SCMC}} & GMM & \textbf{0.00} & 1.22 & 0.05 & \textbf{0.02} & 0.60 & 0.46 & \textbf{0.07} & 0.31 & \textbf{0.02} & 29.92 & \textbf{0.21} \\
& Cosine & 4.24 & 12.31 & 2.08 & 2.27 & 5.46 &7.64 & 3.03 & 2.73 & 2.45 & 44.17 & 3.96 \\
& PLDA & 5.29 & 12.72 & 2.61 & 2.90 & 5.76 & 8.33 & 3.76 & 4.72 & 3.14 & 46.47 & 4.98 \\ \hline

\multirow{3}{*}{\rotatebox[origin=c]{90}{RPS}} & GMM & \textbf{0.00} & \textbf{0.02} & 0.10 & 0.10 & \textbf{0.04} &  2.00 & \textbf{0.01} & 0.92 &\textbf{ 0.00} & 45.18 & 0.73 \\
& Cosine & 3.73 & 3.32 & 5.06 & 4.90 & 6.25 & 10.62 & 9.03 & 17.21 & 3.79 & 46.11 & 10.16\\
& PLDA & 4.20 & 3.74 & 4.46 & 4.12 & 4.49 &  11.11 & 14.38 & 17.03 & 4.53 & 46.93 & 11.76 \\ \hline

\multirow{3}{*}{\rotatebox[origin=c]{90}{MGD}} & GMM & 0.10 & 3.45 & 0.08 & 0.11 & 2.42 & 4.26 & 0.96 & 2.42 & 1.74 & 24.32 & 2.34 \\
& Cosine & 7.19 & 14.74 & 5.04 & 5.48 & 11.42 & 12.42 & 11.82 & 13.00 & 11.09 & 36.59 & 12.08 \\
& PLDA & 8.17 & 15.33 & 4.88 & 5.33 & 11.74 &  13.37 & 13.01 & 13.53 & 11.03 & 38.94 & 12.73 \\ \hline \hline
\multirow{3}{*}{\rotatebox[origin=c]{90}{Fusion1}} & GMM & \textbf{0.00} & \textbf{0.02} & \textbf{0.00} & \textbf{0.00} & \textbf{0.02} & \textbf{0.11} & \textbf{0.04} & \textbf{0.01} & \textbf{0.00} & 21.44 & \textbf{0.04} \\
& Cosine & 0.29 & 1.33 & 0.24 & 0.27 & 1.13 & 2.11 & 0.88 & 1.39 & 0.38 & 41.50 & 1.19 \\
& PLDA & 0.51 & 2.26 & 0.24 & 0.26 & 0.94 & 2.34 & 1.25 & 2.08 & 0.50 & 44.39 & 1.54 \\ \hline
\multirow{3}{*}{\rotatebox[origin=c]{90}{Fusion2}} & GMM & \textbf{0.00} & \textbf{0.00} & \textbf{0.00} & \textbf{0.00} & \textbf{0.00} & 8.36 & 8.48 & 8.61 & 8.35 & \textbf{8.27} & 8.45 \\
& Cosine & 0.96 &  0.91 &  0.87 &  0.92 &  0.91 & 14.26 &  14.41 &  14.53 &  14.26 &  14.45 & 14.36 \\
& PLDA & 0.76 &  0.80 &  0.70 &  0.76 &  0.73 &  15.00 &  15.02 &  15.21 &  14.89 &  15.08 & 15.03 \\ \hline
\noalign{\hrule height 0.75pt}
\end{tabular}
\end{center}
\label{evaluation_set_individual_clean_results}
\end{table}

Independent of the classifier and features, S10 ---the speech synthesis algorithm that uses MARY text-to-speech system\footnote{\url{http://mary.dfki.de/}}--- is the most difficult attack type to detect in comparison to the other unknown attacks (S6-S9).
This could be because S10 does not use any vocoder in generating the synthetic speech signals whereas the popular STRAIGHT vocoder \cite{Straight} is used in most of the remaining attacks.
Thus, spoofing detectors trained with a STRAIGHT vocoder but tested without it will induce a mismatch between the training and the test samples \cite{ASVspoofOverview}, making detection of S10 relatively more difficult.

In general, the SCMC features yield lower EERs than MFCCs with the GMM classifier except for S10.
Concerning the two phase-based features, RPS outperforms MGD in most cases.
Notably, MGD yields considerably better performance than RPS for S10, therefore for unknown attacks, on average.
For the unknown attacks, MFCCs are superior to phase based MGD features.
However, for known attacks RPS yields better accuracy than magnitude based MFCCs.
For the two scoring variants of i-vector, in turn, MFCCs outperform the SCMC features, except for S10.
Overall, S10 yields extremely high EERs while reasonable accuracies are obtained for the other attacks. In most studies that report their findings on the ASVspoof 2015 data, the performance of countermeasures is reported by averaging the EER of individual unknown attacks (S1-S10), which was the official evaluation metric of the challenge. However, the average EER of unknown attacks becomes highly dependent on the performance of S10 attack. Therefore, in Table~\ref{evaluation_set_individual_clean_results}, the performance of unknown conditions are reported by averaging the S6-S9 attacks rather than S6-S10.
Since GMM outperformed i-vectors systematically, only the GMM results are presented in the remaining experiments on the Evaluation set.

Note that in Table 5, simple score averaging (Fusion 1) performs considerably better than fusion with weights optimized using logistic regression (Fusion 2). This stems from the fact that, during the training of Fusion 2, we pool all scores together and look for a joint transformation for all the attack types. This results in almost equal performance of the system to each attack type. Unfortunately, due to a very high EER for S10, this performance could be called as being ``equally bad".

The results for the noise-contaminated evaluation set obtained with GMM using selected magnitude and phase based features are given in Table~\ref{evaluation_set_results}.
MFCCs yield lower EERs than SCMCs under white noise for both known and unknown attacks.
For the babble and car noises, in turn, SCMCs outperform MFCCs.
Similar to results on Development Set (Table~\ref{comp_features_table_ivector}), a considerable reduction in EERs is obtained using SCMC features over MFCCs under car and babble noise cases.
For phase features, RPS is superior to MGD features for both known and unknown attacks under babble and car noises whereas MGD shows better performance than RPS under white noise case.
In general, magnitude features (MFCCs and SCMCs) yield lower EERs than phase features independent of noise and SNR.

\begin{table}
\caption{Comparison (EER, $\%$) of known and unknown attacks for Gaussian Mixture model classifier on evaluation set. In each row, the lowest EERs for the known (K) and unknown (U) attacks (S6-S9 attacks) are bolded and underlined, respectively.}
\begin{center}
\footnotesize
\begin{tabular}{cc||cc||cc||cc||cc||cc|cc}
\noalign{\hrule height 0.75pt}
Noise & SNR & \multicolumn{2}{c||}{MFCC} & \multicolumn{2}{c||}{SCMC} & \multicolumn{2}{c||}{RPS} & \multicolumn{2}{c||}{MGD} & \multicolumn{2}{c|}{Fusion1} & \multicolumn{2}{c}{Fusion2} \\
type & (dB) & K. & U. & K. & U. & K. & U. & K. & U. &K. &  U. & K. &  U. \\ \noalign{\hrule height 0.75pt}
\multicolumn{2}{c||}{Original} & 0.85 & 0.63 & 0.38 & 0.22 & 0.05 & 0.73 & 1.23 & 2.35 & 0.01 & \underline{0.04} & \textbf{0.00} & \underline{0.04} \\
\noalign{\hrule height 0.25pt}
\multirow{3}{*}{\rotatebox[origin=c]{90}{White}} & 20 & 16.43 & 17.94 & 19.92 & 15.40 & 38.53 & 40.62 & 27.25 & 36.24 & \textbf{13.39} & \underline{13.93} & 16.76 & 16.40 \\
& 10 & 25.45 & 29.78 & 33.36 &32.14 & 42.16 & 44.98 & 37.42 & 38.66 & \textbf{22.78} & \underline{26.13} & 25.91 & 27.71\\
& 0 & 35.07 & 39.66 & 43.73 & 42.27 & 44.56 & 46.64 & 44.42 & 45.88 & \textbf{34.29} & \underline{38.53} & 34.96 & 38.90\\
\noalign{\hrule height 0.25pt}
\multirow{3}{*}{\rotatebox[origin=c]{90}{Babble}} & 20 & 7.48 & 6.49 & 2.15 & \underline{1.39} & 6.09 & 10.62 & 14.20 & 23.55 & 1.13 & 1.81 &  \textbf{0.69} & 1.97 \\
& 10 & 15.59 & 12.76 & 8.32 & \underline{5.30} & 21.17 & 23.71 & 26.30 & 35.65 & 5.81 & 6.52 & \textbf{5.36} & 8.08 \\
& 0 & 33.54 & 28.40 & \textbf{29.74} & 25.13 & 40.66 & 40.81 & 37.59 & 40.77 & \textbf{24.90} & \underline{23.75} & 25.23 & 23.95 \\
\noalign{\hrule height 0.25pt}
\multirow{3}{*}{\rotatebox[origin=c]{90}{Car}} & 20 & 3.57 & 2.83 & 0.79 & 0.52 & 0.74 & 3.67 & 9.39 & 16.12 & 0.11 & 0.45 & \textbf{0.05} & \underline{0.38} \\
& 10 & 7.31 & 6.03 & 2.16 & \underline{1.67} & 5.28 &  9.93 & 15.99 & 24.44 & 1.00 & 2.07 & \textbf{0.72} & 1.95 \\
& 0 & 17.33 & 14.69 & 8.59 & \underline{7.36} & 24.66 & 25.67 & 30.32 & 36.63 & 8.17 & 9.38 & \textbf{7.11} & 8.03 \\
\noalign{\hrule height 0.75pt}
\end{tabular}
\end{center}
\label{evaluation_set_results}
\end{table}

\section{Conclusion}
In this study, our goal was to analyze the robustness of existing state-of-the-art countermeasure systems for synthetic speech detection in the presence of additive noise. Extensive experiments were conducted using different front-ends and back-ends for three type of noises (white, babble and car) with three different noise levels (20 dB, 10 dB, and 0 dB). We evaluated the performance with five different short-term magnitude features (MFCC, IMFCC, SCMC, CQCC and MHEC) and three short-term phase features (RPS, MGD, and CosPhase). These features have successfully been used for spoofing detection in clean conditions whereas our study addresses their performance under additive noise backgrounds. As a back-end, we have experimented with two well-known approaches: Gaussian mixture model (GMM) and i-vector. We also explored the effect of various speech enhancement techniques as well as the impact of different feature post-processing methods. Finally, we have investigated fusion techniques to combine the strength of multiple systems.

Our extensive results on ASVspoof 2015 dataset indicate that additive noise contamination considerably complicates the task of synthetic speech detection. Applying standard speech enhancement techniques, such as magnitude spectral subtraction, power spectral subtraction, and Wiener filtering were not found helpful in improving the accuracy. In recent studies, it was reported that DNN-based speech enhancement techniques outperforms standard methods such as MMSE and Wiener filtering \cite{Sun2016}. Therefore, applying DNN-based speech enhancement for anti-spoofing under additive noise would be interesting for the future work. We also found that phase-based features, RPS, and CosPhase, perform better in the absence of any feature post-processing schemes like delta features or cepstral mean subtraction (CMS). But those post-processing steps were found crucial for the other features.

White noise degrades the accuracy the most. For example, in an experiment on the development set, EER increased from $0.84\%$ to $31.42\%$ with MFCC features and GMM back-end in the presence of white noise with 0 dB SNR. The severity of white noise can be explained with the help of comparative long-term average spectra (LTAS) of different noises. We have shown that it has a considerable effect on the entire speech spectrum unlike other two types of noises where the effect on speech spectrum is mostly partial.

Concerning the back-ends, we have observed the GMM-based classifier to consistently outperform the more sophisticated i-vector method. Poor results for the i-vector systems could be explained by short utterances or possibly suboptimal data selection to train UBM and T-matrix. Our findings on spoofing detection task also agree with the results from the previously conducted independent studies, but on the clean condition.

In the study of features using GMM as the classifier, MFCCs give best recognition accuracy in most cases in the presence of white noise while SCMCs perform better for babble and car. However, this observation is not consistent when we take i-vector systems into account. For example in an experiment with the development set, MHEC feature outperforms the other features for the i-vector-cosine system whereas MFCCs win when PLDA scoring is employed. We have also observed that RPS feature --- which was successfully used in many spoofing detection studies and outperforms other features such as MFCC in clean conditions --- generally yield higher EERs than standard MFCC features in the presence of additive noise. However, its performance is still superior to the other two phase based features compared: MGD and CosPhase.

The results on the evaluation section of ASVspoof 2015 further reveals that detecting \emph{unknown} attacks is much harder than detecting \emph{known} attacks in noisy condition. Moreover, from a detailed study on attack-specific performances with clean speech data, we find that the notable performance difference between known and unknown attacks is mostly due to one specific spoofing attach, S10 (i.e., MARY TTS) which does not use any vocoder as the other synthetic speech generation techniques used in ASVspoof 2015. This was the general observation regarding different systems submitted to ASVspoof 2015 \cite{ASVspoofOverview}.

Finally,  we have observed considerable gain in spoofing detection performance due to fusion of multiple front-ends. For example, in the presence of 10 dB car noise, the EERs of known and unknown attack using score-average fused system are $0.99\%$ and $7.82\%$, respectively, whereas best individual system (here, SCMC) gives $2.16\%$ and $8.49\%$. We have also noticed that improvement for the known attack condition is relatively higher than the improvement in unknown attack. We further observe that logistic regression based fusion scheme is better for known attacks, however, score average based method is more appropriate for unknown attacks. This is because for the logistic regression approach the fusion parameters are optimized on development set, i.e., for known attacks, and those optimized parameters are used for fusion of evaluation set scores consisting known and unknown attack. Applying score average based fusion strategy is a compromise to reduce the generalization error. Preventing fusion overfitting is an important practical consideration and clearly deserves further attention.

Our results suggest that synthetic speech detection becomes more challenging in noisy conditions, similar to speaker verification in a noisy environment. This study opens a few potential directions for future work. The first one is a development of robust approaches for both front-end and back-end sides of spoofing detection systems. In front-end side, we used the most promising (or otherwise popular) features in this study. Other phase based techniques, such as RPI that was reported to perform well under noisy conditions in other speech processing tasks, would be interesting to study in antispoofing under additive noise. The other direction is a study of trustworthiness of voice biometric systems under a joint presence of spoofing attacks and noise that calls for joint optimization and evaluation of ASV and spoofing countermeasure systems.

\section{Acknowledgments}
This project has been primarily supported by the Academy of Finland (projects $253120$ and $283256$). The paper also reflects some results from the OCTAVE Project (\#$647850$), funded by the Research European Agency (REA) of the European Commission, in its framework programme Horizon 2020. The views expressed in this paper are those of the authors and do not engage any official position of the European Commission. The work of Cemal Hanil\c{c}i is supported by the Scientific and Technological Research Council of Turkey (T\"{U}B\.{I}TAK) under project $\#$115E916

\bibliographystyle{IEEEtran}
\bibliography{References}

\end{document}